\documentclass[preprint,nohyper]{JHEP3} %
\usepackage{epsfig,multicol}
\usepackage{latexsym}

\usepackage{amssymb,amsmath}

\def\simge{\mathrel{%
   \rlap{\raise 0.511ex \hbox{$>$}}{\lower 0.511ex \hbox{$\sim$}}}}
\def\simle{\mathrel{
   \rlap{\raise 0.511ex \hbox{$<$}}{\lower 0.511ex \hbox{$\sim$}}}}

\newcommand{\newc}{\newcommand}
\newc\eg{{\it {e.g.}}}  \newc\etal{{\it {et al.}}} \newc\ie{{\it i.e.}}
\newc\etc{{\it {etc}}}  

\newcommand\lsim{\mathrel{\rlap{\lower4pt\hbox{\hskip1pt$\sim$}}
    \raise1pt\hbox{$<$}}}
\newcommand\gsim{\mathrel{\rlap{\lower4pt\hbox{\hskip1pt$\sim$}}
    \raise1pt\hbox{$>$}}}

\newc{\mhalf}{m_{1/2}}      \newc{\mzero}{m_0}

\newc{\tanb}{\tan\beta}
\newc{\azero}{A_0}
\newc{\at}{A_t} \newc{\abot}{A_b} \newc{\atau}{A_\tau} 

\newc{\bmu}{B\mu}           \newc{\sgn}{{\rm sgn}}
\newc{\mone}{M_1}           \newc{\mtwo}{M_2}

\newc{\charone}{\chi_1^\pm} \newc{\mcharone}{m_{\chi_1^\pm}}

\newc{\hl}{h}               \newc{\mhl}{m_{\hl}}
\newc{\hh}{H}               \newc{\mhh}{m_{\hh}}
\newc{\ha}{A}               \newc{\mha}{m_{\ha}}
\newc{\hc}{H^{\pm}}         \newc{\mhc}{m_{\hc}}

\newc{\mw}{m_{W}}      \newc{\mz}{m_{Z}}

\newc{\mgut}{M_{\rm GUT}}

\newc{\mplanck}{M_{\rm P}}      \newc{\mpl}{M_{\rm Pl}}
\newc{\msusy}{M_{\rm SUSY}}      \newc{\ms}{M_{\rm S}}

\newc{\jxf}{J({\xf})}
\newc{\jxfexact}{J_{\rm exact}({\xf})}  \newc{\jxfexp}{J_{\rm exp}({\xf})}
\newc{\VEV}[1]{\langle #1 \rangle}

\newc{\xf}{x_f}
\newc\vrel{v_{\rm rel}}
\newcommand\mchi{m_{\chi}}              

\newc\sell{{\widetilde e}_L}      \newc\msell{m_{\sell}}
\newc\selr{{\widetilde e}_R}      \newc\mselr{m_{\selr}}

\newc\snue{{\widetilde \nu}_e}      \newc\msnue{m_{\snue}}
\newc\snutau{{\widetilde \nu}_\tau}      \newc\msnutau{m_{\snutau}}

\newc\supl{{\widetilde u}_L}      \newc\msupl{m_{\supl}}
\newc\supr{{\widetilde u}_R}      \newc\msupr{m_{\supr}}
\newc\sdl{{\widetilde d}_L}      \newc\msdl{m_{\sdl}}
\newc\sdr{{\widetilde d}_R}      \newc\msdr{m_{\sdr}}

\newcommand{\stauone}{{\tilde \tau}_1}   \newcommand\mstauone{m_{\stauone}}
   
\newcommand{\stauleft}{{\tilde \tau}_L}   
\newcommand{\stauright}{{\tilde \tau}_R}

\newcommand\gluino{\tilde g}
\newcommand\mgluino{m_{\gluino}}

\newc\hpm{H^\pm} \newc\hp{H^+} \newc\hm{H^-} 
\newc\sfermion{\tilde f}  \newc\msfermion{m_{\sfermion}}  

\newc\second{{\rm sec}} 

\newc\alphas{\alpha_s}

\newc\alphaem{\alpha_{em}}

\newcommand\treh{T_{\rm R}}

\newc{\sthw}{\sin\theta_W}              \newc{\cthw}{\cos\theta_W}
\newc{\bino}{\widetilde B}              \newc{\wino}{\widetilde W_3}
\newc{\higgsinob}{{\widetilde H}^0_b}   \newc{\higgsinot}{{\widetilde H}^0_t}

\newc{\abund}{\Omega h^2}
\newc{\abundchi}{\Omega_\chi h^2}
\newc{\abundcdm}{\Omega_{{\rm CDM}} h^2}

\newc{\omegam}{\Omega_{{\rm M}}}       \newc{\abundm}{\Omega_{{\rm M}} h^2}
\newc{\omegab}{\Omega_{{\rm b}}}	\newc{\abundb}{\Omega_{{\rm b}} h^2}
\newc{\omegatot}{\Omega_{{\rm TOT}}}

\newc{\omeganlsp}{\Omega_{{\rm NLSP}}}   \newc{\abundnlsp}{\Omega_{\rm NLSP}h^2}
\newc{\ynlsp}{Y_{{\rm NLSP}}}            \newc{\taunlsp}{\tau_{{\rm NLSP}}}
\newc{\nnlsp}{n_{{\rm NLSP}}}            \newc{\mnlsp}{m_{{\rm NLSP}}}
\newc{\nx}{n_{X}}                        \newc{\yx}{Y_{X}}
\newc{\mx}{m_{X}}                        \newc{\taux}{\tau_{X}}

\newc{\rhocrit}{\rho_{crit}}
\newc{\rhochi}{\rho_{\chi}}

\newcommand\fa{f_{a}}

\newcommand\stau{\tilde{\tau}}

\newcommand\neut{\tilde \chi}

\newc{\cachigamma}{C_{a\neut\gamma}}
\newc{\caww}{C_{aWW}}                   
\newc{\cayy}{C_{aYY}}

\newc{\nl}{\cos \theta_{\tilde t}}
\newc{\nr}{\sin \theta_{\tilde t}}
\newcommand\tev{\,\mbox{TeV}}
\newcommand\gev{\,\mbox{GeV}}
\newcommand\mev{\,\mbox{MeV}}
\newcommand\kev{\,\mbox{keV}}
\newcommand\ev{\,\mbox{eV}}

\newc\gbar{{\overline{g}}}

\newc{\ra}{\rightarrow}

\newc{\beq}{\begin{equation}}
\newc{\eeq}{\end{equation}}
\newc{\bea}{\begin{eqnarray}}
\newc{\eea}{\end{eqnarray}}

\newc{\nspin}{n_{\rm spin}}
\newc{\nflavor}{n_{\rm F}}
\newc{\ngamma}{n_\gamma}
\newc{\ychi}{Y_{\chi}}                  \newc{\yeqchi}{Y^{\rm EQ}_{\chi}}

\newcommand\axino{\tilde{a}}

\newc{\naxino}{n_{\axino}}
\newc{\yaxino}{Y_{\axino}}
\newc{\yeqaxino}{Y^{\rm EQ}_{\axino}}
\newc{\ythaxino}{Y^{\rm TP}_{\axino}}
\newc{\ynthaxino}{Y^{\rm NTP}_{\axino}}

\newcommand\gravitino{\widetilde{G}}    
\newcommand\mgravitino{m_{\gravitino}}

\newcommand\abundg{\Omega_{\gravitino}h^2}
\newcommand\abundgntp{\Omega^{\rm NTP}_{\gravitino}h^2}     

\newcommand\abundgtp{\Omega^{\rm TP}_{\gravitino}h^2}       

\newc{\ngravitino}{n_{\gravitino}}
\newc{\ygravitino}{Y_{\gravitino}}
\newc{\yeqgravitino}{Y^{\rm EQ}_{\gravitino}}
\newc{\ythgravitino}{Y^{\rm TP}_{\gravitino}}
\newc{\ynthgravitino}{Y^{\rm NTP}_{\gravitino}}

\newc{\yascat}{Y^{\rm scat}_{i,j}}      \newc{\yadec}{Y^{\rm dec}_{i}}
\newc{\gstar}{g_\ast}           \newc{\gsstar}{g_{s\ast}}

       \def\pslash{\not{\hbox{\kern-2.3pt $p$}}}
       \def\kslash{\not{\hbox{\kern-2.3pt $k$}}}
       \def\qslash{\not{\hbox{\kern-2.3pt $q$}}}
       \def\ddslash{\not{\hbox{\kern-2.3pt $d$}}}
       \def\prtslash{\not{\hbox{\kern-2.3pt $\partial$}}}

\newcommand\jcap[3] 
		{{\it J.\ Cosmol.\ Astrop.\ Phys.\ }{\bf #1} (#2) #3}

\title{Gravitino Dark Matter in the CMSSM\\
  With Improved Constraints from BBN}
\author{David G. Cerde\~no\\
Institute for Particle Physics Phenomenology, University of
Durham, DH1 3LE, UK\\
E-mail: \email{d.g.cerdeno@durham.ac.uk}
}
\author{Ki-Young Choi\\
Department of Physics and Astronomy, University of Sheffield, 
Sheffield, S3 7RH, UK\\
E-mail: \email{K.Choi@sheffield.ac.uk}
}
\author{Karsten Jedamzik\\
Laboratoire de Physique Th\'eorique et Astroparticules,
CNRS UMR 5825,\\
Universit\'e Montpellier II, F-34095 Montpellier Cedex 5, France\\
E-mail: \email{jedamzik@LPM.univ-montp2.fr}
}
\author{Leszek Roszkowski\\
Department of Physics and Astronomy,
University of Sheffield, Sheffield, S3 7RH, UK\\
E-mail: \email{L.Roszkowski@sheffield.ac.uk}
}
\author{Roberto Ruiz de Austri\\
Departamento de F\'{\i}sica Te\'{o}rica C-XI
 and Instituto de F\'{\i}sica Te\'{o}rica C-XVI,
 Universidad Aut\'{o}noma de Madrid, Cantoblanco,
 28049 Madrid, Spain\\
E-mail: \email{rruiz@delta.ft.uam.es}
}

\abstract{
In the framework of the Constrained MSSM we re--examine the gravitino
  as the lightest superpartner and a candidate for cold dark matter in
  the Universe. Unlike in most of other recent studies, we include
  both a thermal contribution to its relic population from scatterings
  in the plasma and a non--thermal one from neutralino or stau decays
  after freeze--out.  Relative to a previous analysis~\cite{rrc04} we
  update, extend and considerably improve our treatment of constraints
  from observed light element abundances on additional energy released
  during BBN in association with late gravitino production. Assuming
  the gravitino mass $\mgravitino$ in the $\gev$ to $\tev$ range, and
  for natural ranges of other supersymmetric parameters, the
  neutralino region is excluded, except for rather exceptional cases, while for smaller values of
  $\mgravitino$ it becomes allowed again.  The gravitino relic
  abundance is consistent with observational constraints on cold dark
  matter from BBN and CMB in some well defined domains of the stau
  region but, in most cases, only due to a dominant contribution of
  the thermal population. This implies, depending on $\mgravitino$, a
  large enough reheating temperature. If $\mgravitino>1\gev$ then
 $\treh>10^7\gev$, if allowed by BBN and other constraints but, for
  light gravitinos, if $\mgravitino>100\kev$ then $\treh>  10^3\gev$.
  On the other hand, constraints mostly from BBN imply an
  upper bound 
$\treh\lsim \textrm{a few}\times10^8\gev$ 
  which appears inconsistent 
with thermal leptogenesis.
  Finally, most of the preferred
  stau region corresponds to the physical vacuum being a false
  vacuum. The scenario can be partially probed at the LHC.
}
\preprint{}

\begin{document}

\section{Introduction}\label{sect:intro}

Weakly interacting massive particles (WIMPs) remain the most popular
choice for cold dark matter (CDM) in the Universe. This is because
they are often present in various extensions of the Standard Model
(SM). For example, 
neutralinos in softly broken low energy SUSY models can be made
stable by assuming some additional symmetries (like $R$--parity in
SUSY). Their relic abundance in some regions of the
parameter space agrees with the value of $\abundcdm\sim0.1$ inferred
from observations. This last property is sometimes
taken as a hint for a deeper link between electroweak physics and the
cosmology of the early Universe.

However, extremely weakly interacting massive particles
(E--WIMPs),\footnote{Another name used in the literature is
  `superweakly interacting massive particles'~\cite{feng03-prl}.}
have also been known to provide the desired values of the CDM relic
density. E--WIMPs are particles whose interactions with ordinary
matter are strongly suppressed compared to ``proper'' WIMPs, like
massive neutrinos and neutralinos, whose interactions are set by the
SM weak interaction strength $\sigma_{weak}\sim
10^{-38}{\,\mbox{cm}^2}$ times some factors, like mixing angles for
the neutralino which are often much smaller than one. For such WIMP
their physics is effectively entirely determined by the Fermi scale
and (in the case of SUSY) by the SUSY breaking scale $\msusy$, which,
for the sake of naturalness, is expected not to significantly exceed
the electroweak scale.

In contrast, a typical interaction strength of E--WIMPs is suppressed
by some large mass scale $m_\Lambda$, $\sigma_{\rm
E-WIMP}\sim(\mw/m_\Lambda)^2\sigma_{weak}$.  One particularly
well--known example is the gravitino for which $m_\Lambda$ is the
(reduced) Planck scale $\mplanck=1/\sqrt{8\pi
G_N}=2.4\times10^{18}\gev$. Another well--motivated possibility is
axions and/or their fermionic partner axinos for which the scale
$m_\Lambda$ is given by the Peccei--Quinn scale
$\fa\sim10^{11}\gev$. Both the axion~\cite{axiondm} and the
axino~\cite{ckrplus} have also been shown to be excellent candidates
for CDM. Other possibilities involve moduli~\cite{moduli-cdm} although
they strongly depend on a SUSY breaking mechanism. Thus the
electroweak scale may have little to do with the DM problem, after
all.

In particular, the gravitino as a supersymmetric partner of the
graviton, is present in schemes in which gravity is incorporated into
supersymmetry (SUSY) via local SUSY, or supergravity. As a spin--$3/2$
fermion, it acquires its mass through the super--Higgs mechanism.
Since gravitino interactions with ordinary matter are strongly
suppressed, it was realized early on that the particle was facing
various cosmological problems~\cite{pp82,weinberg-grav82}.  If the
gravitino is not the lightest superpartner (LSP), it decays into the
LSP rather late ($\sim10^{2-10}\sec$) and associated electromagnetic
(EM) or hadronic (HAD) radiation. If too much energy is dumped into
the expanding plasma at late times $\gsim1\sec$, then the associated
particles produced during the decay (\eg, a photon in the gravitino
decay to the neutralino) can cause unacceptable alterations of the
successful predictions of the abundances of light elements produced
during Big Bang nucleosynthesis (BBN), for which there is a good
agreement between theory on the one hand and direct observations and
CMB determinations on the other.  Since the number density of
gravitinos is directly proportional to the reheating temperature
$\treh$, this leads to an upper bound of
$\treh<10^{6-8}\gev$~\cite{khlopov+line84,ekn84,ens84,nos83,jss85,km94}
(for recent updates see, \eg,~\cite{cefo02,kkm04,kmy05}). On the other
hand, when the gravitino is the LSP and stable (the case considered in
this work), ordinary sparticles will first cascade decay into the
lightest ordinary superpartner, which would be the next--to--lightest
superpartner, (NLSP) which would then decay into the gravitino and
associated EM and/or HAD radiation. A combination of this and the
overclosure argument ($\abundg<1$) has in this case led to a rough
upper bound $\treh\lsim 10^9\gev$~\cite{ens84,mmy93}.

There are some generic ways through which gravitinos (assumed from now
on to be the LSP) can be produced in the early Universe. One mechanism
has just been described above: the NLSP
first freezes out and then, at much later times, decays into the
gravitino. Such a process does not depend on the previous thermal
history of the Universe (so long as the freeze--out temperature is
lower than the reheating temperature $\treh$ after inflation). As in
our previous work, we will call it a mechanism of non--thermal
production (NTP). In a class of thermal production (TP) processes
gravitinos can also be generated through scattering and decay
processes of ordinary (s)particles during the thermal expansion of the
Universe. Once produced, gravitinos will not participate in a reverse
process because of their exceedingly weak interactions. Analogous
mechanisms exist in the axino case~\cite{ckrplus}. In addition, there
are other possible ways of populating the Universe with stable relics,
\eg\ via inflaton decay or during preheating~\cite{grt99,kklp99}, or
from decays of moduli fields~\cite{kyy04}. In some of these cases the
gravitino production is independent of reheating temperature and its
abundance may give the measured dark matter abundance with no ensuing
limit on $\treh$. In general, such processes are, however, much more
model dependent and not necessarily efficient~\cite{nps01}, and will
not be considered here.

Recently, there has been renewed interest in the gravitino as a stable
relic and a dominant component of {\em cold} dark matter. Thermal
production was re--considered in~\cite{bbp98,bbb00} while non--thermal
production
in~\cite{feng03-prl,feng03,fst04,eoss03-grav}. In~\cite{fiy03} both
processes were considered in the framework of the Minimal
Supersymmetric Standard Model (MSSM) in the context of thermal
leptogenesis.  In~\cite{rrc04} some of us considered a combined impact
of both production mechanisms in the more predictive framework of the
Constrained MSSM (CMSSM)~\cite{kkrw94}. The CMSSM encompasses a class
of unified models where at the GUT scale gaugino soft masses unify to
$\mhalf$ and scalar ones unify to $\mzero$.  We concentrated on
$\mgravitino$ in the $\gev$ to $\tev$ range, typical of
gravity--mediated SUSY breaking, and on TP contributions at large
$\treh\sim 10^9\gev$.

Since all the NLSP particles decay after freeze--out, in NTP the
gravitino relic abundance $\abundgntp$ is related to $\Omega_{\rm
NLSP} h^2$ -- the relic abundance that the NLSP would have had if it
had remained stable -- via a simple mass ratio
\beq
\abundgntp=\frac{\mgravitino}{\mnlsp}\Omega_{\rm NLSP} h^2.
\label{abundgntp:eq}
\eeq
Note that $\abundgntp$ grows with the mass of the gravitino
$\mgravitino$.

The gravitino relic abundance generated in TP can be
computed by integrating the Boltzmann equation from $\treh$ down to
today's temperature. In the case of the gravitino, a simple formula
for $\abundgtp$ has been obtained in~\cite{bbp98,bbb00}
%
\begin{equation}
\abundgtp\simeq 0.27 \left(\frac{\treh}{10^{10}\gev}\right)
\left(\frac{100\gev}{\mgravitino}\right) 
\left(\frac{\mgluino(\mu)}{1\tev}\right)^2,
\label{eq:abundgbbb}
\end{equation}
where $\mgluino(\mu)$ above is the running gluino
mass. In~\cite{bbp98,bbb00} it was argued that, for natural ranges of
the gluino and the gravitino masses, one can have $\abundgtp\sim 0.1$
at $\treh$ as high as $10^{9-10}\gev$.  

The problem is that in many unified SUSY models, the number density of
stable relics undergoing freeze--out is actually often too large. For
example, in the CMSSM, the relic abundance of the lightest neutralino
typically exceeds the allowed range, except in relatively narrow
regions of the parameter space. This can be easily remedied if the
neutralino is not the true LSP and can decay further, for example into
the gravitino (or the axino~\cite{ckrplus}).  Indeed, it is sufficient
to take a small enough mass ratio in the formula~(\ref{abundgntp:eq})
above. Then, however, $\abundgtp$ may become too large because of its
inverse dependence on $\mgravitino$, especially at high values of
$\treh$, essential for thermal leptogenesis~\cite{fy,crv96}, and for
$\mgravitino$ in the $\gev$ to $\tev$ range.

One may want to suppress the contribution from TP by considering
$\treh\ll10^9\gev$ and generate the desired relic density of
gravitinos predominantly through NLSP freeze--out and decay. This
would normally require a larger gravitino mass $\mgravitino$ and
therefore longer decay lifetimes (see below).  This, however, can lead
to serious problems with BBN, as discussed above.  Furthermore, late
injection of energetic photons into the plasma may distort the nearly
perfect blackbody shape of the CMB spectrum~\cite{hu93}.

In a previous paper~\cite{rrc04} by some of us, the issue of a
combined impact of TP and NTP mechanisms of gravitino production, in
view of requiring the total gravitino
$\abundg=\abundgntp+\abundgtp\sim0.1$ and of BBN, CMB and other
constraints, has been examined in the framework of the
CMSSM.

In the CMSSM, the NLSP is typically either the (bino--dominated)
neutralino (for $\mhalf\ll\mzero$) or the lighter stau $\stauone$ (for
$\mhalf\gg\mzero$). Assuming natural ranges of $\mhalf,\mzero\lsim$~a
few TeV, the whole neutralino NLSP region was found~\cite{rrc04} to be
excluded by constraints from BBN because of unacceptably large showers
generated by NLSP decays even assuming fairly conservative abundances
of light elements. This confirmed the findings of~\cite{fiy03,fst04}.
On the other hand, the fraction of the stau NLSP 
region excluded by the BBN constraint was found to depend rather
sensitively on assumed ranges of abundances of light elements.

However, it was also found in~\cite{rrc04} that the stau NLSP region
of parameter space where $\abundchi\sim0.1$ was due to NTP alone were
in most cases excluded by (mostly) the BBN constraints.  In other
words, for natural ranges of $\mhalf$, a significant component of
$\abundg$ from TP (and thus rather high $\treh$) must normally be
included in studies of gravitino CDM in the CMSSM.

In both the neutralino and the stau NLSP cases, decay products lead
mostly to EM showers but a non--negligible fraction of them develop
HAD showers which can also be very dangerous, especially at shorter
NLSP lifetimes ($\lsim10^4\sec$).  Constraints on EM showers were
analyzed in~\cite{cefo02} assuming rather conservative ranges for the
abundances of light elements but important constraints from HAD showers
were not included.  A combined analysis of constraints on both EM
and HAD fluxes was recently performed in~\cite{kkm04} with much more
restrictive (and arguably in some cases perhaps too restrictive)
observational constraints than~\cite{cefo02}. In~\cite{rrc04} the
EM shower constraint was applied following~\cite{cefo02} and the HAD
one following~\cite{kkm04} and thus, out of necessity, using different
assumptions about allowed ranges of light elements.

In the present analysis we make a number of improvements. Firstly, we
treat both EM and HAD showers in a self--consistent way by assuming the same
ranges of abundances of light elements which we take to be somewhat
less restrictive than those adopted in~\cite{kkm04}. Secondly, the BBN
yields are computed with a sophisticated code which simultaneously
deals with the impact of both EM and HAD showers. In some parts of the parameter space 
this is essential since 
non--linearities may exist and
in such cases EM and HAD activities cannot be analyzed separately in
computing the abundances of light elements. Thirdly, at each point in
the CMSSM we compute energy released into all relevant (EM and HAD)
channels and their hadronic branching ratios (which are typically
smaller than the EM ones but can vary by a few orders of magnitude)
and then use them as inputs into the BBN code.  Furthermore,
in~\cite{rrc04} in dealing with EM showers we conservatively only
included bounds from $D/H$, $Y_p$ (${^4}{\! He}$ abundance) and
${^7}{\! Li}/{H}$ while in constraining HAD showers we dropped the
lithium constraint. In the current work we include all the above three
constraints and in addition apply constraints from ${^3}{\! He}/D$ and
${^6}{\!  Li}/{^7}{\! Li}$ which in some cases have the strongest
impact on the CMSSM parameter space. All these improvements lead to a
much more reliable BBN constraint on EM/HAD showers, even though a large
part of the stau NLSP region still remains allowed. In particular, we
improve the upper bound of $\treh\lsim5\times10^9\gev$ found
in~\cite{rrc04} to $\treh\lsim \textrm{a few}\times10^8\gev$.

In addition, we have now extended the range of $\mhalf$ to 6\tev, beyond that
considered in~\cite{rrc04}, and found that at very large values
of $\mhalf\gsim4\tev$  one can find allowed regions consistent with $\abundg$ in
the interesting range due to NTP alone. We have also found relatively
confined pockets of the neutralino NLSP parameter space which are
consistent with BBN and CMB. We discuss these results below.

Lastly, in SUSY theories the presence of several scalar fields
carrying color and electric charge allows for a possible existence of
dangerous charge and color breaking (CCB) minima~\cite{revmunoz} --
\cite{ccb-cosmology} 
which would render the physical (Fermi) vacuum
unstable.  Along some directions in field space the (tree--level)
potential can also become unbounded from below (UFB).  Avoiding these
instabilities leads to constraints on the parameter space among which
those derived from requiring the absence of UFB directions are by far
the most restrictive~\cite{clm1}. In the specific cases applicable to
the CMSSM, after including one--loop corrections, such UFB directions
become bounded but develop deep CCB minima at large field values away
from the Fermi vacuum of our Universe.

Although the existence of such a dangerous global vacuum cannot be
excluded if the lifetime of the (metastable) Fermi vacuum is longer
than the age of the Universe (in addition to having rather unpleasant
scatological consequences), such a possibility does place non--trivial
constraints on inflationary cosmology. One has to explain why and how
the Universe eventually ended up in the (local) Fermi
minimum~\cite{ccb-cosmology,forss96}. 

The effect of the UFB constraints in minimal supergravity models was
analyzed in~\cite{cggm03-1}, where it was shown that it is the 
stau region in the CMSSM
that is mostly affected. This is precisely the region of interest for
gravitino CDM in the CMSSM.  Consequently, we will discuss the impact
of these constraints in our analysis.

As in~\cite{rrc04}, we will take $\mgravitino$ as a free parameter and
allow it to vary over a wide range of values from ${\cal O}(\tev)$
down to the sub--$\mev$ range, for which the gravitino (at least those
produced in TP, see later) would remain {\em cold} DM relic.  Lighter
gravitinos would become warm and then (sub--$\kev$) hot DM.  We will
not address the question of an underlying (if any) supergravity model
and SUSY breaking mechanism. As in~\cite{rrc04}, we will mostly focus
on the ${\cal O}(\gev)$ to ${\cal O}(\tev)$ mass range, as most
natural in the CMSSM with gravity--mediated SUSY breaking, but will
also explore at some level light gravitinos.

In the following, we will first summarize our procedures for computing
$\abundg$ via both TP and NTP. Then we will list NLSP decay modes into
gravitinos, and discuss constraints on the CMSSM parameter space, in
particular those from BBN and CMB. Finally, we will discuss
implications of our results for thermal leptogenesis and for SUSY
searches at the LHC.

\section{Framework and Procedure}

Unless otherwise stated, we will follow the analysis and notation of
the previous paper~\cite{rrc04} to which we refer the reader for more
details. Here we only summarize the main points, while below we
elaborate on our improved analysis of constraints from BBN and on a
previously neglected impact of CCB minima and UFB directions.

Within the framework of the CMSSM we employ two--loop RGEs to compute
both dimensionless quantities (gauge and Yukawa coupling) and
dimensionful ones (gaugino and scalar masses) at the electroweak
scale. Mass spectra of the CMSSM are determined in terms of the usual
five free parameters: the previously mentioned $\tanb$, $\mhalf$ and
$\mzero$, as well as the trilinear soft scalar coupling $\azero$ and
$\sgn(\mu)$ -- the sign of the supersymmetric Higgs/higgsino mass
parameter $\mu$.  The parameter $\mu$ is derived from the condition of
electroweak symmetry breaking and we take $\mu>0$. We compute the mass
spectra with the help of the package SUSPECT~v.~2.34
~\cite{suspect:ref}.  For simplicity we assume $R$--parity
conservation even though E--WIMPs, like gravitinos or axinos, can
constitute CDM even when it is broken. This is because, even in the
presence of $R$--parity breaking interactions the E--WIMP lifetime
will normally be very large due to their exceedingly tiny interactions
with ordinary matter.

We compute the number density of the NLSP after freeze--out (the
neutralino or the stau) with high accuracy by numerically solving the
Boltzmann equation including all (dominant and subdominant) NLSP pair
annihilation and coannihilation channels. For a given value of
$\mgravitino$, we  then compute the NTP contribution to the
gravitino relic abundance $\abundgntp$ via
eq.~(\ref{abundgntp:eq}). In computing the thermal contribution
$\abundgtp$ we employ~eq.~(\ref{eq:abundgbbb}).

\begin{figure}[!t]
  \begin{center}
  \begin{tabular}{c c}
    \includegraphics[width=0.5\textwidth]{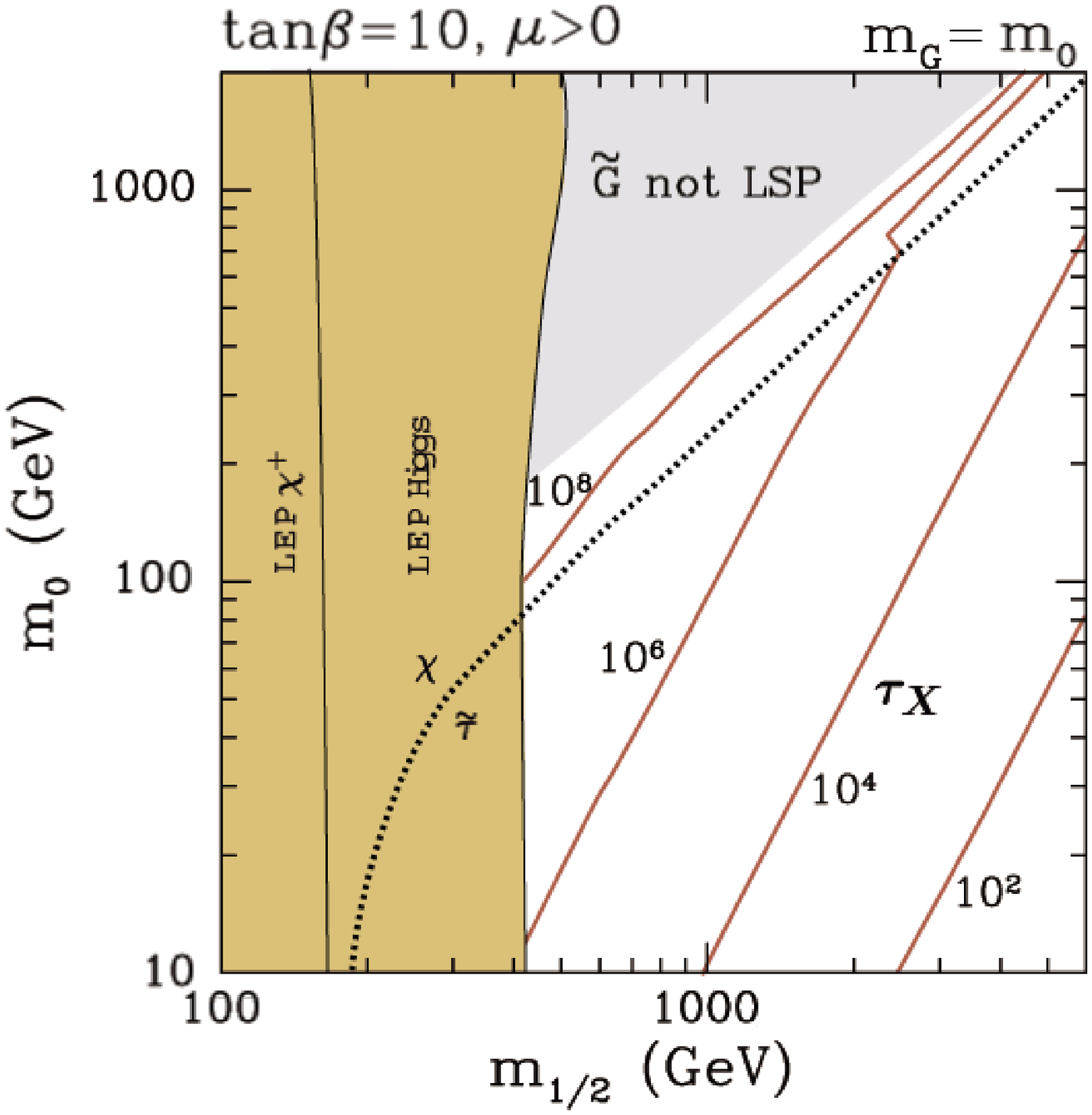}
    & 
   \includegraphics[width=0.5\textwidth]{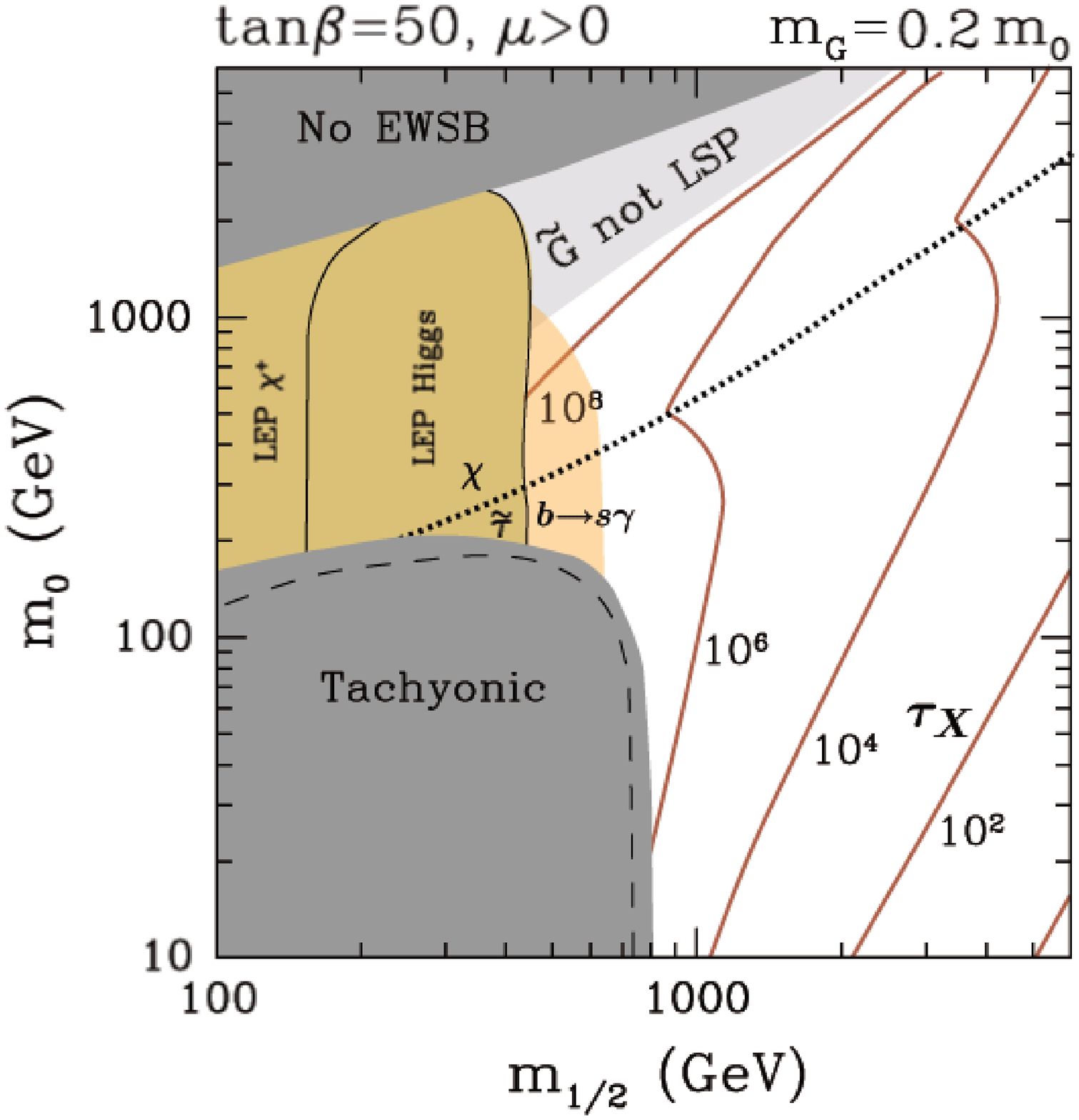}
  \end{tabular}
  \end{center}
  \caption{\small The plane ($\mhalf,\mzero$) for $\tanb=10$,
$\mgravitino=\mzero$ (left window) and $\tanb=50$,
$\mgravitino=0.2\mzero$ (right window) and for $\azero=0$, $\mu>0$. 
The light brown regions labelled ``LEP $\chi^+$''
and ``LEP Higgs'' are excluded by unsuccessful chargino and Higgs
searches at LEP, respectively. In the right window the darker brown
region labelled ``$b\to s\gamma$'' is excluded assuming minimal flavor
violation. The dark grey region below the dashed line is labelled
``TACHYONIC'' because of some sfermion masses becoming tachyonic and
is also excluded. In the rest of the grey region (above the dashed
line) the stau mass bound $\mstauone>87\gev$ is violated. In the region
``No EWSB'' the conditions of EWSB are not satisfied. 
Magenta lines mark contours of the NLSP lifetime $\taux$ (in seconds).
The dotted line is the boundary of neutralino ($\chi$) or stau ($\stau$) 
NLSP. 
}
\label{fig:lifetime}
\end{figure}

After freeze--out from the thermal plasma at $t\sim10^{-12}\sec$, the NLSPs
decay into gravitinos at late times which strongly depend on the NLSP
composition and mass, on $\mgravitino$ and on the final states of the
NLSP decay. Expressions for $\Gamma_X=1/\taux$, where $X$ denotes the
decaying particle\footnote{From now on we will 
denote $X=\chi,\stauone$ for brevity.}  have been derived
in~\cite{eoss03-grav,fst04}. Given some discrepancies between
the two sources, below (as in~\cite{rrc04}) we
follow~\cite{fst04}.  Roughly, for $\mgravitino\ll \mx$ the lifetime is given by
\beq
\taux\sim
10^8\sec\left(\frac{100\gev}{\mx}\right)^5
\left(\frac{\mgravitino}{100\gev}\right)^2.
\label{lifetime:eq}
\eeq
The exact value of the NLSP lifetime in the
CMSSM further depends on a possible relation between $\mgravitino$ and
$\mhalf$ and/or $\mzero$ but in the parameter space allowed by other
constraints it can vary from $\gsim10^8\sec$ at smaller $\mx$ down
to $10^2\sec$, or even less, for large $\mhalf$ and/or $\mzero$ in the
$\tev$ range.

In fig.~\ref{fig:lifetime} we present the gravitino relic abundance
and the NLSP lifetime in the usual plane spanned by $\mhalf$ and
$\mzero$ for two representative choices: $\tanb=10$ and
$\mgravitino=\mzero$ (left window) and $\tanb=50$ and
$\mgravitino=0.2\mzero$ (right window), and for $\azero=0$ and
$\mu>0$. 
At small $\mzero$ and large
$\tanb$ some sfermions become tachyonic, as encircled by a dashed line
inside the grey region (labelled ``TACHYONIC'') in the right
window. Relevant collider and theoretical constraints (but not yet
those coming from BBN or CMB) are shown.  We apply the same
experimental bounds as in~\cite{rrc04}: (i) the lightest chargino mass
$\mcharone >104\gev$, (ii) the lightest Higgs mass $m_h>114.4\gev$,
(iii) $BR(B\rightarrow X_s\gamma) = (3.34 \pm 0.68)\times10^{-4}$. In
addition, now we further impose a stau mass bound
$\mstauone>87\gev$~\cite{sleptons} which slightly enlarges the grey
region beyond the part labelled ``TACHYONIC''.  In this analysis we
update the top quark mass to the current value of
$m_t=172.7\gev$~\cite{top172}.

To help understanding
this and subsequent figures, we remind the reader of some basic mass
relations. The mass of the gluino is roughly given by $\mgluino\simeq
2.7\mhalf$.  The mass of the lightest neutralino, which in the CMSSM
is almost a pure bino, is $\mchi\simeq0.4\mhalf$. The lightest stau
$\stauone$ is dominated by $\stauright$ and well above $\mz$ its mass
is (neglecting Yukawa contributions at large $\tanb$) roughly given by
$\mstauone^2\simeq \mzero^2+0.15\mhalf^2$.  This is why at
$\mzero\ll\mhalf$ the stau is lighter than the neutralino while in the
other case the opposite is true. The boundary between the two NLSP
regions is marked with a roughly diagonal dotted line. (In the
standard scenario the region of a stable, electrically charged stau
relic is thought to be ruled out on astrophysical grounds.) Regions
corresponding to the lightest chargino and Higgs masses below their
LEP limits are appropriately marked and excluded. Separately marked
for $\tanb=50$ is the region inconsistent with the measured branching
ratio $BR(B\rightarrow X_s\gamma)$. (For $\tanb=10$, and generally not
too large $\tanb$, this constraint is much weaker and ``hides''
underneath the above LEP bounds.) In the grey wedge of large $\mzero$
conditions of EWSB cannot be satisfied. Finally, for some combinations of
parameters the gravitino is not the LSP. We exclude such cases in this
analysis.


Also shown in fig.~\ref{fig:lifetime} are contours of the NLSP
lifetime $\taux$ (in seconds).  Their shape strongly depends on
gravitino mass relation with $\mzero$, $\mhalf$ and, because of SUSY
mass relations, on other parameters which determine $\mx$, but in the
cases considered here, at small to moderate $\mzero$, $\taux$ typically
decreases with increasing $\mhalf$.

Finally, by comparing fig.~\ref{fig:lifetime} with fig.~2
of~\cite{rrc04} (where $m_t=178\gev$ was assumed), we can see the
sensitivity to the top quark mass.  The region disallowed by the LEP
Higgs mass bound has now broadened from $\mhalf\sim 300\gev$ to $400
\gev$ and at low $\mhalf$ and large $\mzero$ a wedge inconsistent with
EWSB has appeared. There is also a noticeable change is in the
pseudoscalar resonance region for $\tan\beta=50$. The green band moves
to smaller $\mhalf$ and higher $\mzero$ for smaller $m_t$.

\section{Improved BBN Analysis}

NLSP ($\chi$ or $\stauone$) decays after freeze--out can generate highly
energetic electromagnetic and hadronic fluxes which can
significantly alter the abundances of light elements. 
At longer lifetimes $\taux\gsim10^4\sec$ EM constraints are strongest
but at earlier times HAD shower constraint typically become dominant
while the EM one virtually disappears. However, even at later
times HAD constraints can be important.

We will evaluate the abundances of light elements produced during BBN
in the presence of EM/HAD showers and, by comparing them with
observations, place bounds on the latter.  To this end, we need to know the energy
$\epsilon^X_i$ transferred to each decay channel $i=em,had$ and their
respective branching fractions to EM/HAD showers $B^X_i$ as well as
the NLSP lifetime $\taux$.  All the above quantities depend on the
NLSP and (with the exception of the yield) on its decay modes and the
gravitino mass. For the cases of interest ($\chi$ and $\stauone$)
these have been recently evaluated in detail
in~\cite{fst04}. In~\cite{rrc04} and below we follow their discussion.

For the neutralino NLSP the dominant decay mode is
$\chi\ra\gravitino\gamma $. In the CMSSM the
neutralino is a nearly pure bino, thus $\chi\simeq \widetilde{B}$.
The decay $\chi\ra\gravitino\gamma$ produces mostly EM energy.  Thus 
\begin{eqnarray}
\epsilon^\chi_{em}  = \frac{\mchi^2-\mgravitino^2}{2\mchi}, \qquad \qquad
B^\chi_{em} \simeq  1.
\label{brchiem:eq}
\end{eqnarray}

Above their respective kinematic thresholds, the neutralino can also decay via
$\chi\ra \gravitino Z, \gravitino h, \gravitino H, \gravitino A$ for
which the decay rates are given in~\cite{fst04,eoss03-grav}.
These processes contribute to HAD fluxes because of large
hadronic branching ratios of the $Z$ and the Higgs bosons
($B^Z_{had}\simeq 0.7$, $B^h_{had}\simeq 0.9$). 
In this case the transferred energy  $\epsilon^X_i$ and 
branching fraction $B^X_i$ each channel are 
\begin{eqnarray}
\epsilon^\chi_{k} \approx 
\frac{\mchi^2-m_{\gravitino}^2+m_{k}^2}{2\mchi},\qquad
B^\chi_{had,\,k} = 
\frac{\Gamma(\chi\ra\gravitino k)B^k_{had}}{\Gamma_{tot}},
\qquad k=Z,h, H, A,
\end{eqnarray}
where
\begin{eqnarray}
\Gamma_{tot}\simeq\Gamma\left(\chi\ra\gravitino\gamma \right) + 
\sum_k \Gamma\left(\chi\ra\gravitino k \right).
\end{eqnarray}

Below the kinematic threshold for neutralino decays into $\gravitino$
and the $Z$/Higgs boson, one needs to include 3--body decays with the
off--shell photon or $Z$ decaying into quarks for which
$\epsilon^\chi_{q\bar{q}} \approx \frac{2}{3}(m_\chi-m_{\tilde{G}})$
and $B^\chi_{had}(\chi\ra\gravitino \gamma^\ast/Z^\ast \ra \gravitino
q\bar q)\sim 10^{-3} $~\cite{fst04}. This provides a lower bound on
$B^\chi_{had}$.  At larger $\mchi$, Higgs boson final states become
open and we include neutralino decays into them as well.

The dominant decay mode of the lighter stau $\stauone$ is
$\stauone\ra\gravitino\tau$ which, as argued in~\cite{feng03,fst04},
contributes basically only to EM showers. Thus
\begin{eqnarray}
\\
\epsilon^{\stauone}_{em} \approx \frac{1}{2}
\frac{\mstauone^2-\mgravitino^2}{2\mstauone},\qquad\qquad
B^{\stauone}_{em} \simeq  1, 
\label{brstauem:eq}
\end{eqnarray}
where the additional factor of $1/2$ appears because about half of the
energy carried away by the tau--lepton is transmitted to final state
neutrinos. Our results are not sensitive to an order of two variations
of this overall pre--factor.

As shown in~\cite{fst04}, for stau NLSP, the leading contribution
to HAD showers come from 3--body decays
${\stauone}\ra\gravitino\tau Z, \gravitino\nu_{\tau}W$, or from 4--body
decays ${\stauone}\ra\gravitino\tau \gamma^\ast/Z^\ast \ra
\gravitino\tau q \bar q$. 
The transferred energy  $\epsilon^X_i$ and 
branching fractions $B^X_{had,\,i}$ ($i=Z,W,q\bar{q}$) are 
 
\begin{eqnarray}
\epsilon^{\stauone}_{Z}
\simeq\epsilon^{\stauone}_{W}\simeq\epsilon^{\stauone}_{q\bar{q}}
\approx \frac{1}{3}(m_{\stauone}-m_{\gravitino})
\label{epsstauhad:eq}
\end{eqnarray}
and
\begin{eqnarray}
B^{\stauone}_{had,\,Z} =\frac{\Gamma(\stauone\ra\gravitino \tau
  Z)B^Z_{had}}{\Gamma_{tot}},\ \ 
B^{\stauone}_{had,\,W} =\frac{\Gamma(\stauone\ra\gravitino \nu_\tau
  W)B^W_{had}}{\Gamma_{tot}},\ \ 
B^{\stauone}_{had,\,{q\bar{q}}} =\frac{\Gamma(\stauone\ra\gravitino \tau
  q\bar q )}{\Gamma_{tot}}, 
\end{eqnarray}
where
\begin{eqnarray}
\Gamma_{tot}\simeq\Gamma\left(\stauone\ra\gravitino\tau
\right)+\Gamma(\stauone\ra\gravitino \tau  Z)+
\Gamma(\stauone\ra\gravitino \nu_\tau W). 
\end{eqnarray}
One typically finds $B^{\stauone}_{had}\sim10^{-5}- 10^{-2}$ when
3--body decays are allowed and $\sim10^{-6}$ from 4--body decays
otherwise, thus providing a lower limit on the
quantity~\cite{fst04}. (Since the process $\stauone\ra\gravitino
\nu_\tau W$ in proportional to the $\stauleft$ component of
$\stauone$, which in the CMSSM is suppressed, its contribution to
$\Gamma_{tot}$ is likely to be tiny.) Given such a large variation in
$B^{\stauone}_{had}$, the choice~(\ref{epsstauhad:eq}) is probably as
good as any other.

For each point in the parameter space and for a given $\mgravitino$,
the partial energies $\epsilon_i^X$ released into all the channels and
their respective branching fractions $B^{X}_{i}$ are passed on to the
BBN code which computes light element abundances in the presence of
additional EM/HAD showers. A determination of $\epsilon_i^X$ for all
the individual hadronic channels is necessary since the changes of
light element yields are not simply a linear function of
$\epsilon_i^X$. (Note that in our previous analysis~\cite{rrc04} this
yield dependence on $\epsilon_i^X$ was neglected.)  The output is then
compared with observational constraints.

We emphasize that our treatment of the branching
ratios constitutes a significant improvement relative to previous
analyses where only a few sample calculations at a fixed NLSP
mass (\eg, 100\gev\ or  1\tev) and for only a fixed hadronic branching ratio
(\eg, $B_{had}= 10^-3$) were linearly extrapolated to derive BBN yield
predictions.  This is inaccurate for several reasons. Hadronic BBN
yields are not simply linear functions of the hadronic energy
injected, but rather depend in a more complicated way on the energy of
the hadronic primaries. Furthermore, at early times ($\tau\lsim 10^4\sec$) a
simple linear extrapolation from one calculation with particular $B_{had}$ is
impossible due to the interplay between the hadronic perturbations and
thermal nuclear reactions. Finally, at later times, cancellation
effects between HAD and EM light element
production and destruction processes may occur. All these effects may
be properly addressed only when for each point in the SUSY parameter space
an separate BBN calculation is performed.

At early times $\tau \lsim 10^4\sec$ limits from BBN on particle decay
induced showers come from injections of hadrons, \ie, mesons for
$\tau \lsim 10^2\sec$ and nucleons for $\tau \gsim 10^2\sec$, with
EM showers having no effect at such early times. Mesons
convert protons to neutrons by charge exchange reactions, e.g.  $\pi^-
+ p\to \pi^0 + n$~\cite{renoseckel88}, thereby increasing the final
${^4}{\! He}$ abundance.  Nucleons lead to an increase in the $D$
abundance due to both injected neutrons fusing to form $D$ or
inducing the spallation of ${^4}{\! He}$ and concomitant production
of $D$~\cite{dehs88+89}.  Injected energetic nucleons may also affect
a very efficient ${^6}{\! Li}$ production for $\tau\gsim 10^3$
sec~\cite{dehs88+89,jedamzik04,kkm04}. At times $\tau \gsim 10^4$ sec
HAD showers are still important in setting constraints, unless
$B^X_{had}$ is very small. In addition, EM showers, also
lead to distortions of the light element abundances by
photo disintegrating elements~\cite{ens84}. EM showers
typically lead to elevated ${^3}{\! He}/D$~\cite{sjsb95,kkm04} and
${^6}{\! Li}/{^7}{\!  Li}$~\cite{jedamzik99} ratios.  (Note that the
effects of HAD showers have not been considered in
ref.~\cite{eoss03-grav}).

In the present analysis we fix the baryon--to--photon ratio $\eta$ at
$6.05\times 10^{-10}$ which is consistent with the WMAP result $\eta =
6.1^{+0.3}_{-0.2} \times 10^{-10}$~\cite{wmap_cdm}. All processes
required for an accurate determination of the light  element
abundances are treated in detail.  The calculations are based on the
code introduced in ref.~\cite{jedamzik04} with the effects of
EM showers added. Details of this code will be presented
elsewhere~\cite{jedamzik-bbncode}. (A similar detailed presentation
can be found in ref.~\cite{kkm04}.) 

We  apply the following observational constraints
\begin{center}
\begin{tabular}{r c l}
$2.2\times10^{-5}<$ & $D/H$ & $< 5.3\times10^{-5}$ 
\\ 
$0.232 <$ & $Y_p$ & $< 0.258$ 
\\
$8\times10^{-11}<$ & ${^7}{\! Li}/H$ & 
\\

& ${^3}{\! He}/D$ & $< 1.72$ 
\\
& ${^6}{\! Li}/{^7}{\! Li}$ & $< 0.1875$.
\end{tabular}
\label{lightelements:table} 
\end{center}

Note that in~\cite{kkm04} a much less conservative upper bound
of $3.66\times10^{-5}$ on $D/H$ was assumed. On the other hand, 
the constraint from ${^6}{\!  Li}/{^7}{\! Li}$ was not applied.  Note
also that, relative to~\cite{rrc04}, we now also include
important constraints from ${^3}{\! He}/D$ and ${^6}{\! Li}/{^7}{\!
Li}$. These bounds are typically a factor of ten more stringent than
constraints from $D$ alone. Though the stellar evolution of ${^3}{\!
He}$ is not well understood, the constraint from the ${^3}{\! He}/D$
ratio is very secure as $D$ is known to always be destroyed in stars
whereas ${^3}{\! He}$ may be either destroyed or produced in
stars. Thus one may derive an upper limit on the primordial ${^3}{\!
He}/D$ ratio given by its value in the pre--solar nebula. Less secure
is the upper limit on the ${^6}{\! Li}/{^7}{\!  Li}$ ratio, as
${^6}{\! Li}$ is more fragile than ${^7}{\!  Li}$. Typical ${^6}{\!
Li}/{^7}{\!  Li}$ observations in low--metallicity stars fall in the
range $\sim 0.03 - 0.1$ (cf.~\cite{lambert04}).

\begin{figure}[!t]
  \begin{center}
  \begin{tabular}{c c}
   \includegraphics[width=0.5\textwidth]{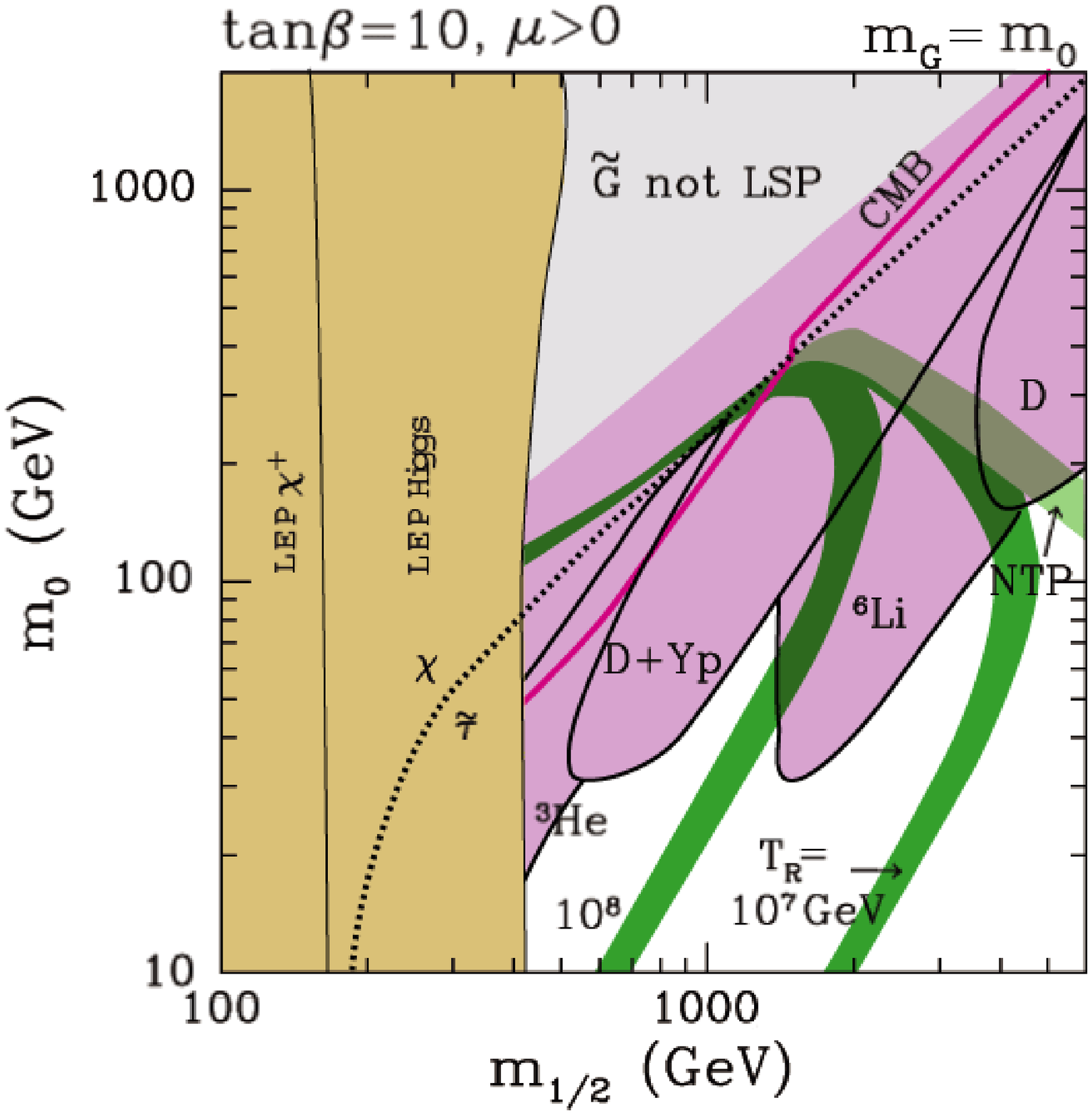}
    & 
   \includegraphics[width=0.5\textwidth]{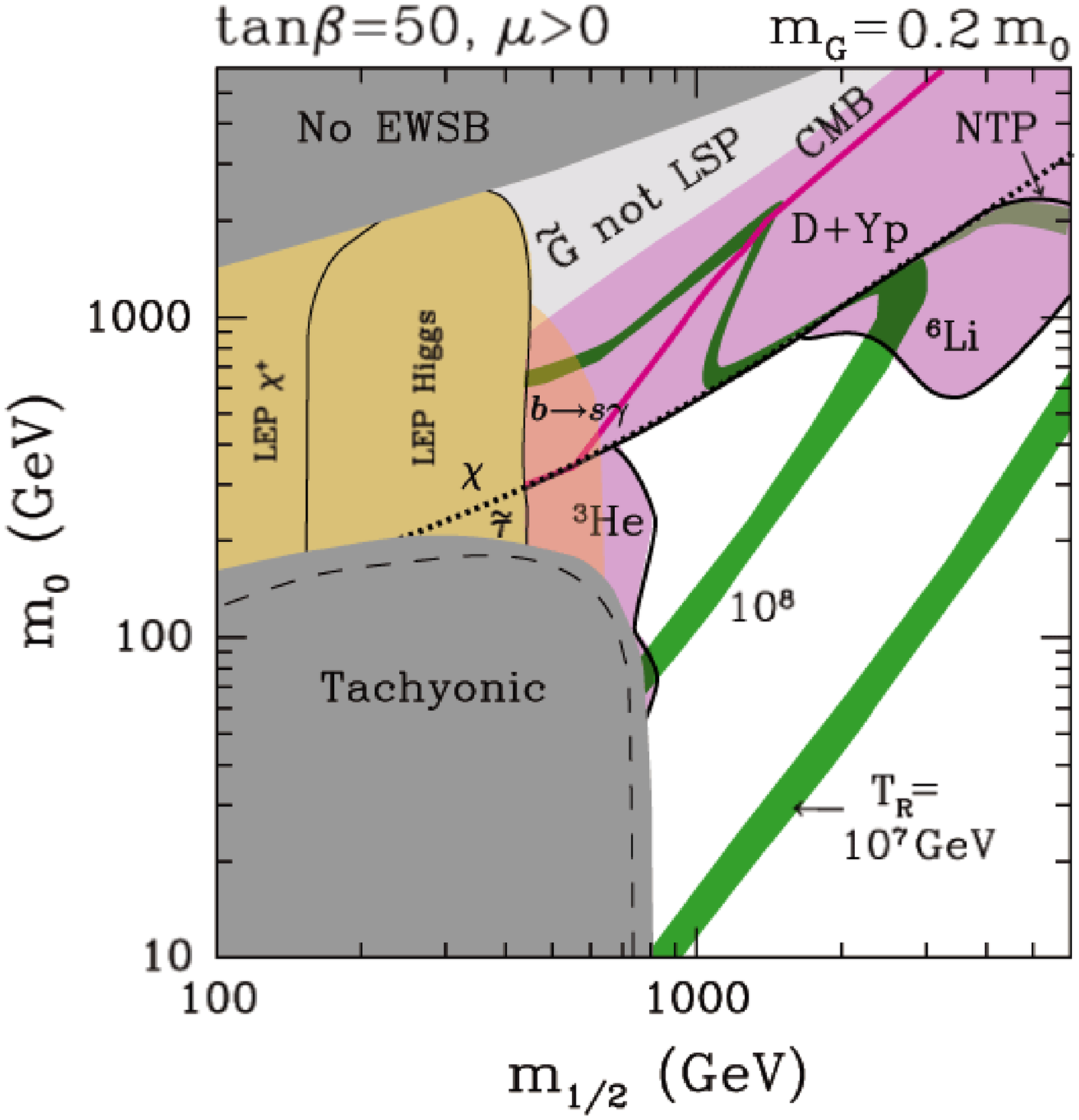}
  \end{tabular}
  \end{center}
  \caption{\small The same as in fig.~\protect\ref{fig:lifetime} but
  with constraints from BBN and CMB superimposed. The regions excluded
  by the various BBN constraint are denoted in violet.  The region
  disallowed by $D+Y_p$ and additional regions excluded by ${^3}{\!
  He}$ and ${^6}{\! Li}$ are denoted accordingly by their respective
  names. A solid magenta curve labelled ``CMB'' delineates the region
  (on the side of label) inconsistent with the CMB spectrum.
In both  windows, 
 the dark green bands labelled ``$T_R=10^7 \gev$''  and ``$10^8$''
 denote the total relic abundance of the gravitino from both thermal 
 and non--thermal production with denoted reheating temperature 
 is in the favored range, while in the light green regions
 (marked ``NTP'') the same is the case for the relic abundance from NTP
 processes alone.}
  \label{fig:newbbn}
\end{figure}

Finally we mention the constraint from the CMB shape. 
Late injection of electromagnetic energy may distort the frequency
dependence of the CMB spectrum from its observed blackbody
shape~\cite{ekn84,hu93}, as recently re--emphasized
in~\cite{feng03} and in~\cite{rrc04}. In this paper we
apply our analysis presented in~\cite{rrc04} to which refer the reader
for all the details.

In fig.~\ref{fig:newbbn} we present in the usual ($\mhalf,\mzero$)
plane our new and more accurate constraints from BBN for the same
parameters as in fig.~\ref{fig:lifetime}.  First we note that a robust
constraint from $D/H$ and $Y_p$ excludes basically the whole
neutralino NLSP region.\footnote{However, we do find some limited
  exception to this, as we discuss below.} This remains true so long as
$\mgravitino>{\cal O}(1\gev)$ as we will discuss later. This is
consistent with the previous analysis~\cite{rrc04} and also confirms the
findings of~\cite{fiy03,fst04}.  Next, the important role played by
the constraints from ${^3}{\! He}/D$ and ${^6}{\!  Li}/{^7}{\! Li}$ in
excluding additional regions of the ($\mhalf,\mzero$) is shown
explicitly.

The cases presented in fig.~\ref{fig:newbbn} correspond to some of the cases
presented in fig.~2 in ref.~\cite{rrc04} in order to allow a
comparison with a previous approximate treatment of the BBN
constraint. We can see a substantial weakening of the HAD shower
bound coming from $D/H$. This is because  in ref.~\cite{rrc04} a much
stronger upper bound on the allowed abundance of deuterium was
applied, following ref.~\cite{kkm04}.  Otherwise, the general pattern
of excluded regions is roughly similar.

The total gravitino relic abundance consistent with the $2\,\sigma$
range $0.094< \abundcdm < 0.129$~\cite{wmap_cdm} (marked $\abundg$) is
shown 
in dark green. For comparison, light
green regions (marked NTP) correspond to the gravitino relic abundance
due to NTP alone in the same range. These regions become
cosmologically favored when one does not include TP or when $\treh\ll
10^9\gev$.  (Note that these regions correspond to ranges of $\mhalf$
beyond those explored in~\cite{eoss03-grav}, where only NTP was
considered, and were not found there.)  In the white regions encircled
by the dark (light) green bands, the total (NTP--induced) relic
abundance is too small while on the other side it is too large.  Note
that the shape of the green bands strongly depends on the gravitino
mass relation with $\mzero$, $\mhalf$ and/or other
parameters~\cite{rrc04}.  As emphasized in~\cite{rrc04}, at large
$\treh$ these two bands of $\abundg$ and $\abundgntp$ correspond to
very different regions of stau NLSP parameter space.

On the other hand, it is only at such large
$\mhalf$, where the BBN constraint becomes much weaker due to a much
shorter lifetime, that we find some cases where NTP alone can be
efficient enough to become consistent with preferred range of CDM
abundance. This can be seen in the left window of
fig.~\ref{fig:newbbn}.

We also note that the constraint from not distorting the CMB
spectrum (magenta line) seems generally less important than that due
to BBN~\cite{ld05}.

It is thus clear that, so long as $\treh\lsim \textrm{a few}
10^8\gev$,
one finds sizable
regions of rather large $\mhalf$ and much smaller $\mzero$ consistent
with the preferred range of CDM abundance. Unless one allows for
very large $\mhalf\gsim4\tev$, a substantial (and, in fact, dominant) TP contribution to
$\abundg$ are required.

 \section{False Vacuua}

A complete analysis of all the potentially dangerous CCB and UFB
directions in the field space of the CMSSM, including the radiative
corrections to the scalar potential in a proper way, was carried out
in ref.~\cite{clm1}.  As we commented in the Introduction, the most
restrictive bounds are from the UFB directions, and therefore we will concentrate
on them below.

In the CMSSM there are three UFB directions, labelled in~\cite{clm1} as UFB--1,
UFB--2 and UFB--3. It is worth mentioning here that the unboundedness is
only true at tree level since radiative corrections eventually raise
the potential for large enough values of the fields. Still these
minima can be deeper than the usual Fermi vacuum and thus dangerous.
The UFB--3 direction involves the scalar fields
$\{H_u,\nu_{L_i},e_{L_j},e_{R_j}\}$ with $i \neq j$ and thus leads
also to electric charge breaking. Since it yields the strongest bound
among all the UFB and CCB constraints, and for future convenience, let
us briefly give the explicit form of this constraint.

By simple analytical minimization of the relevant terms of the
scalar potential it is possible to write the
value of all the {$\nu_{L_i},e_{L_j},e_{R_j}$} 
fields in
terms of $H_u$. Then, for any value of $|H_u|<M_{GUT}$ satisfying
\begin{eqnarray}
|H_u| > \sqrt{ \frac{\mu^2}{4\lambda_{e_j}^2}
+ \frac{4m_{L_i}^2}{g'^2+g_2^2}}-\frac{|\mu|}{2\lambda_{e_j}} \ ,
\label{SU6}
\end{eqnarray}
the potential along the UFB--3 direction is simply given
by
\begin{eqnarray}
V_{\rm UFB-3}=(m_{H_u}^2
+ m_{L_i}^2 )|H_u|^2
+ \frac{|\mu|}{\lambda_{e_j}} ( m_{L_j}^2+m_{e_j}^2+m_{L_i}^2 ) |H_u|
-\frac{2m_{L_i}^4}{g'^2+g_2^2} \ .
\label{SU8}
\end{eqnarray}
Otherwise
\begin{eqnarray}
\label{SU9}
V_{\rm UFB-3}= m_{H_u}^2
|H_u|^2
+ \frac{|\mu|} {\lambda_{e_j}} ( m_{L_j}^2+m_{e_j}^2 ) |H_u| + \frac{1}{8}
(g'^2+g_2^2)\left[ |H_u|^2+\frac{|\mu|}{\lambda_{e_j}}|H_u|\right]^2 \ .
\end{eqnarray}
In eqs.~(\ref{SU8}) and~(\ref{SU9}) $\lambda_{e_j}$ denotes the leptonic Yukawa
coupling of the $j$th generation. 
Then, the
UFB--3 condition reads
\begin{eqnarray}
\label{SU7}
V_{\rm UFB-3}(Q=\hat Q) > V_{\rm Fermi} \ ,
\label{conditionufb3}
\end{eqnarray}
where $V_{\rm Fermi}=-\frac{1}{8}\left(g'^2 + g_2^2\right)
\left(v_u^2-v_d^2\right)^2$, with $v_{u,d}=\langle H_{u,d}^0\rangle$,
is the Fermi minimum evaluated at the typical scale of SUSY masses.
(Normally, a good choice for $M_{SUSY}$ is a geometric average of the stop
masses.) The minimization scale $\hat Q$ is given by 
$\hat Q\sim {\rm max}(\lambda_{top} |H_u|, \msusy)$.  With these
choices for $\hat Q$ and $M_{SUSY}$ the effect of the one--loop
corrections to the scalar potential is minimized.
\begin{figure}[!t]
  \begin{center}
  \begin{tabular}{c c}
   \includegraphics[width=0.5\textwidth]{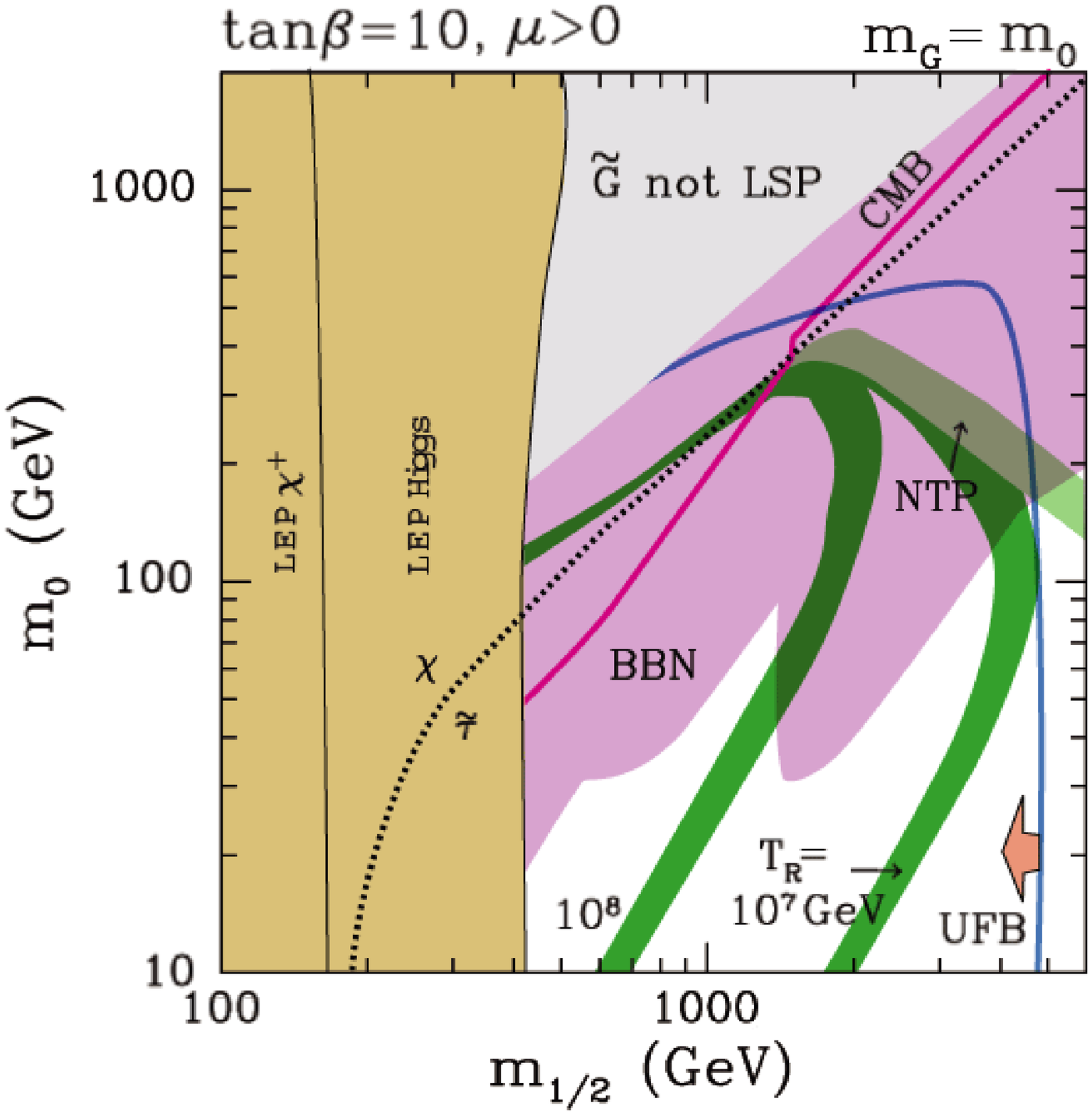}
    & 
  \includegraphics[width=0.5\textwidth]{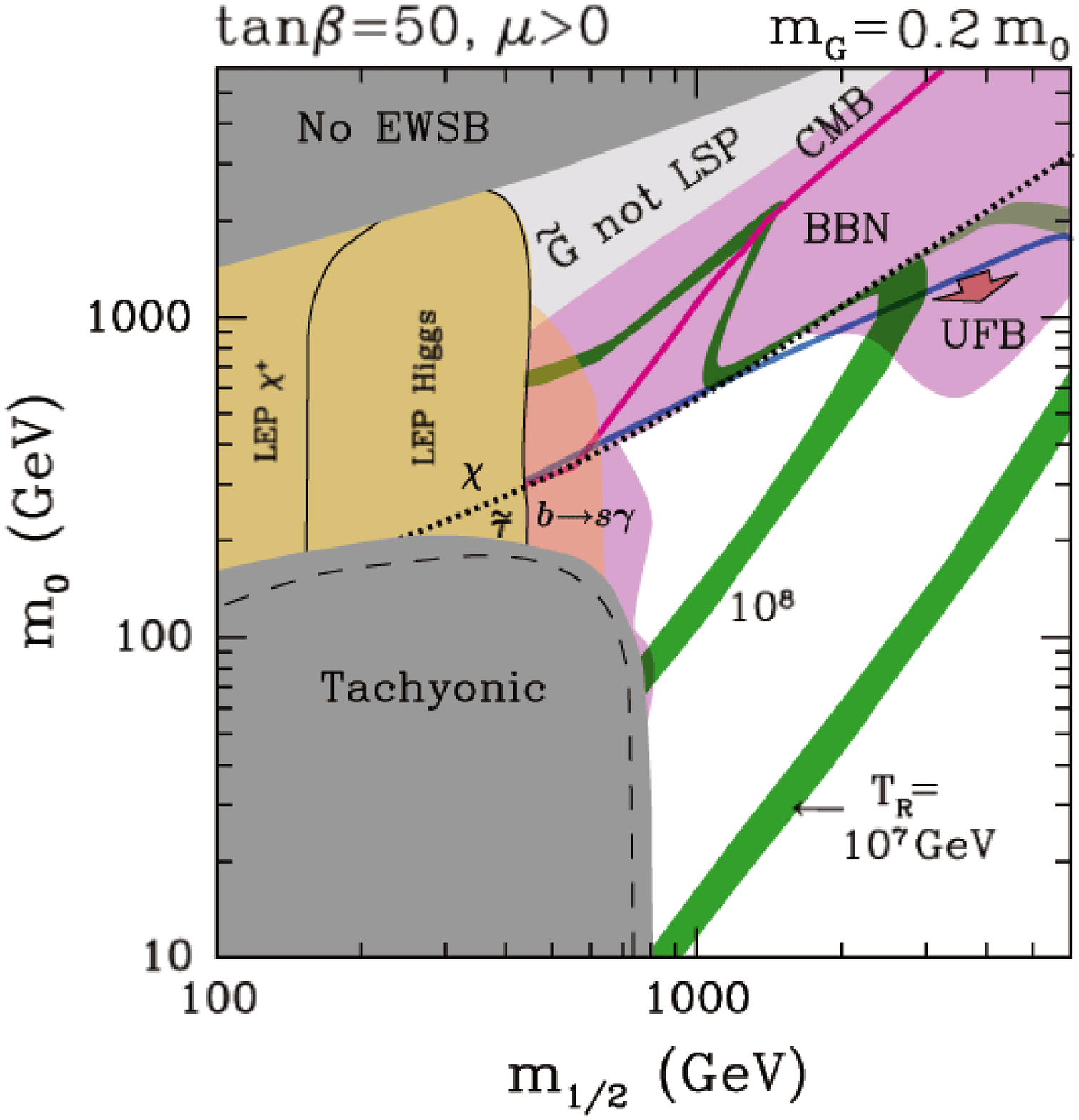}
  \end{tabular}
  \end{center}
  \caption{The same as fig.~\protect\ref{fig:newbbn} but with UFB
  constraints (solid blue line and UFB label plus a big arrow)
  added. For $\mhalf\lsim5\tev$ and small $\mzero$ the UFB constraints
  disfavor the stau NLSP region that 
  has remained allowed after applying the BBN and CMB constraints.}
  \label{fig:bbn+ufb}
\end{figure}
Notice from eqs.~(\ref{SU8}) and~(\ref{SU9}) that the negative
contribution to $V_{\rm UFB-3}$ is essentially given by the $m_{H_u}^2$
term, which in many cases can be large. On the other hand, the
positive contribution is dominated by the term $\propto
1/\lambda_{e_j}$, thus the larger $\lambda_{e_j}$ the more restrictive
the constraint becomes. Consequently, the optimum choice of the
$e$--type slepton is the third generation one, i.e.  ${e_j}=\tilde\tau$.

\begin{figure}[!t]
  \begin{center}
  \begin{tabular}{c c}
   \includegraphics[width=0.5\textwidth]{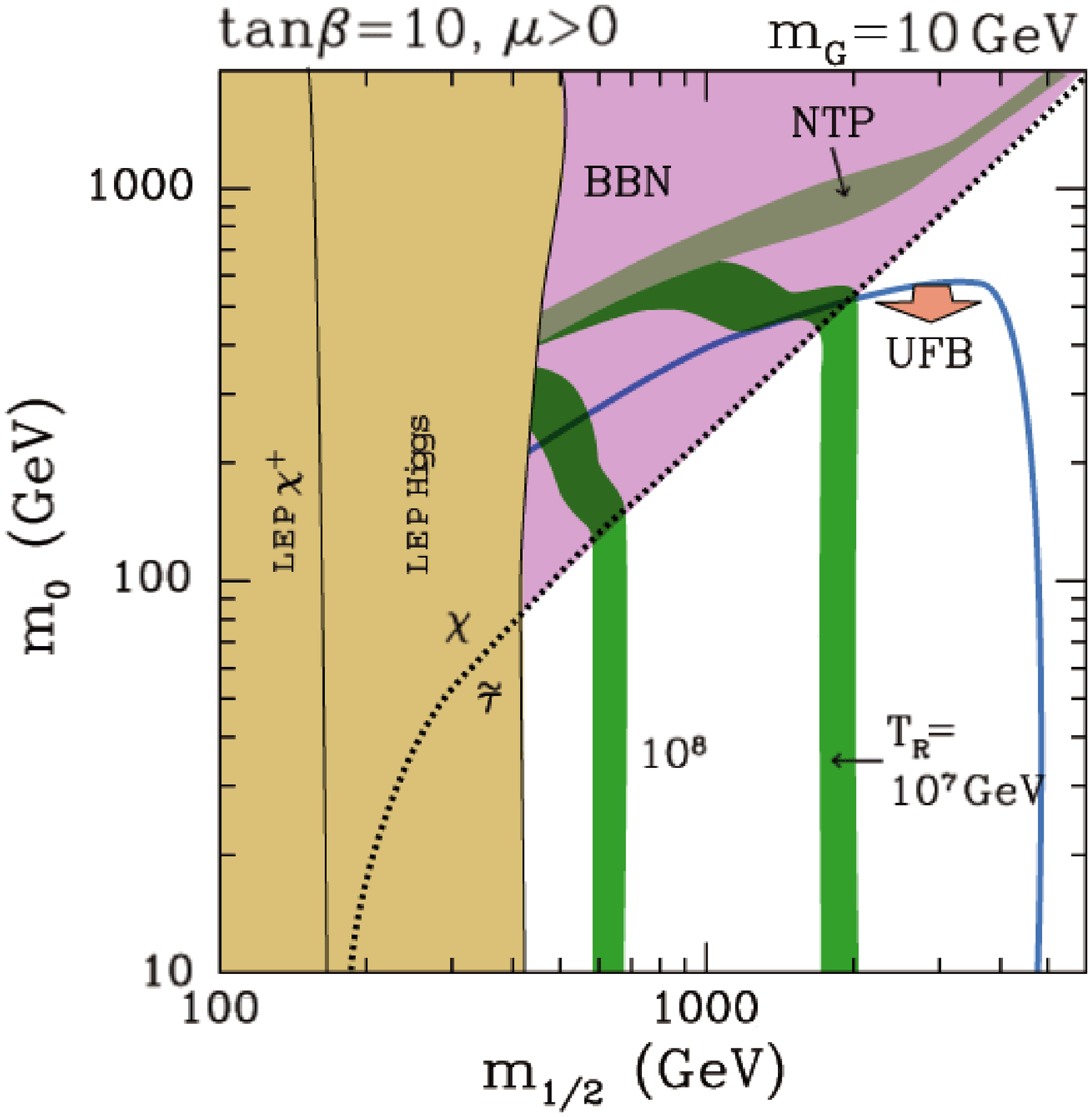}
    & 
   \includegraphics[width=0.5\textwidth]{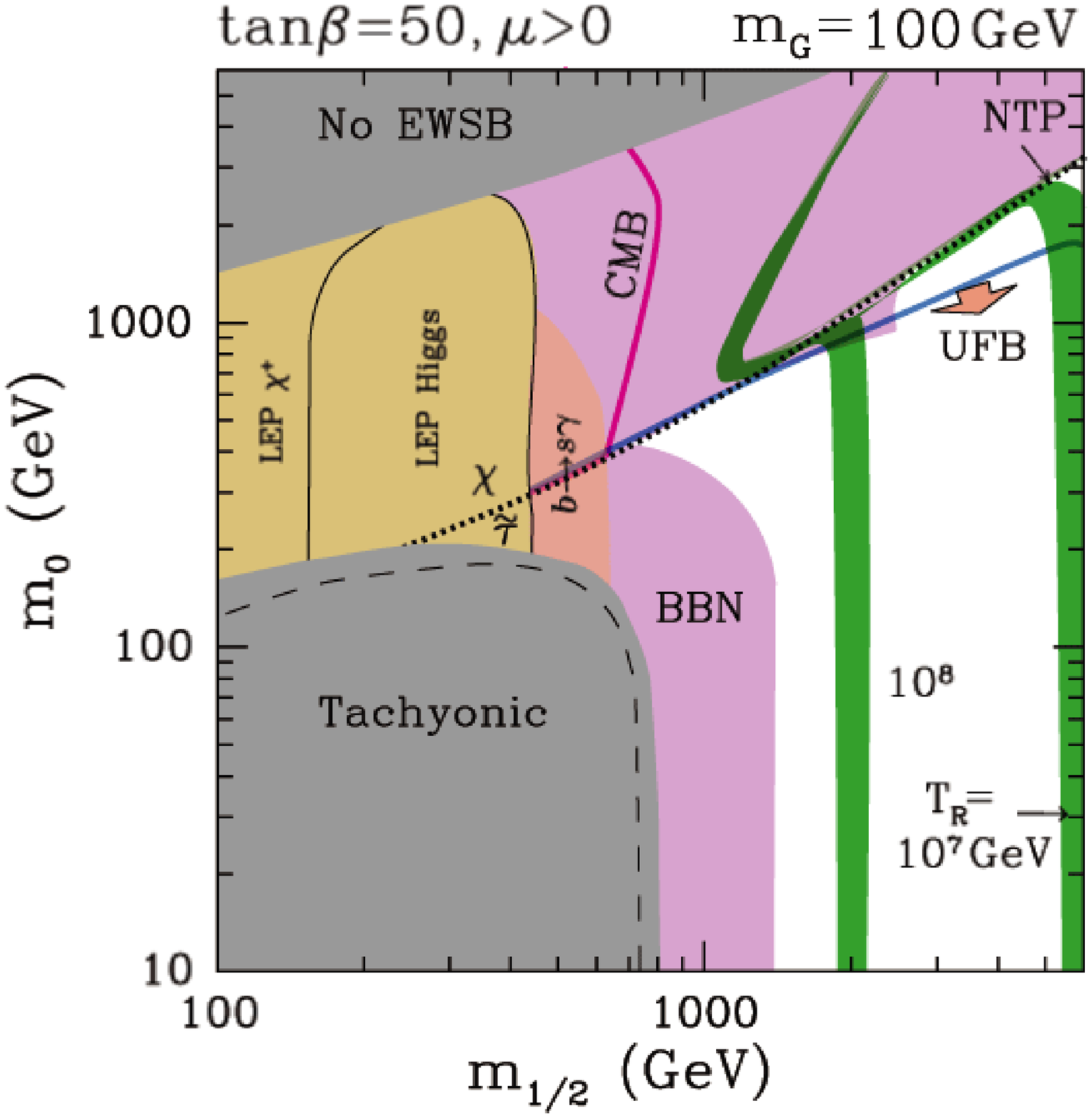}
  \end{tabular}
  \end{center}
  \caption{The same as fig.~\protect\ref{fig:bbn+ufb} but for fixed
    gravitino mass, 
    $\mgravitino=10\gev$ and $\tanb=10$ (left window) and
    $\mgravitino=100\gev$ and $\tanb=50$ (right window).  }
  \label{fig:fixedmG}
\end{figure}
Moreover, since the positive contribution to $V_{\rm UFB-3}$ is
proportional to $m_{\tilde\tau}^2$, the potential will be deeper in
those regions of the parameter space where the staus are light and the
condition~(\ref{conditionufb3}) is more likely to be violated.  For
this reason, the cases with stau NLSP are typically more affected by
the UFB constraints~\cite{cggm03-1}.

In fig.~\ref{fig:bbn+ufb} we present the cases displayed above in
fig.~\ref{fig:newbbn} but now in addition we mark the regions
corresponding to a one--loop corrected UFB--3 direction becoming the
global CCB minimum. These are encircled by a solid blue line and
marked ``UFB'' and a big arrow. We can see that, unless $\mhalf$ is
excessively large ($\mhalf\gsim4\tev$ for $\tanb=10$), in both cases
the whole previously allowed (white) and cosmologically favored
(green) regions correspond to a false vacuum. As $\mstauone$ grows
with increasing $\mhalf$, the UFB constraint becomes weaker and
eventually disappears.

As stated in the Introduction, one cannot exclude the possibility that
the color and electric charge neutral (Fermi) vacuum that the Universe
ended up in after inflation is not a global one but merely a
long--lived local minimum. As discussed in~\cite{fors95,forss96}, this
however often puts a significant constraint on models of cosmic
inflation. The point is that, at the end of inflation, the Universe was
very likely to end up in the domain of attraction of the global
minimum which, even at high temperatures ($\treh\sim10^{7-9}\gev$),
could well have been a CCB one. Preventing such situations leads also to
constraints on the SUSY parameter space.

For this reason, while the UFB regions presented in fig.~\ref{fig:bbn+ufb}
(which correspond to the Fermi vacuum being a local minimum) cannot be
firmly excluded, should SUSY searches at the LHC find sparticles with
masses indicating such a false vacuum, valuable information may be
gained about the state of the Universe and about early Universe
cosmology.

In fig.~\ref{fig:fixedmG} we present two cases with a fixed
$\mgravitino$. In the left window $\tanb=10$ and $\mgravitino=10\gev$
while in the right one $\tanb=50$ and $\mgravitino=100\gev$. (The
``cross'' case of $\tanb=10$ and $\mgravitino=100\gev$ is excluded by
a combination of collider and BBN constraints.)  While now BBN
constraints have become somewhat weaker due to smaller $\mgravitino$
(and therefore smaller $\taux$), 
the whole neutralino NLSP region is
again ruled out as well as part of the stau NLSP region corresponding
to smaller $\mhalf$ (and therefore smaller $\mstauone$ and hence
larger $\taux$). Most of the remaining stau NLSP region in both
windows corresponds to the Fermi vacuum being a local minimum.

As  mentioned earlier, for one case ($\tanb=50$
and $\mgravitino=10\gev$, not shown here) we have found a confined pocket in the neutralino region
around $\mhalf\simeq2.5\tev$ and $\mzero\simeq5\tev$, close to the
region of no EWSB, which is allowed by our BBN and CMB
constraints. However, we consider it to be an odd exception to the
rule rather than a typical case.

Finally, in figs.~\ref{fig:bbn+ufb} and~\ref{fig:fixedmG} for
$\treh=10^8\gev$ 
one finds 
$\mhalf\lsim2\tev$ in order to remain consistent with
$\abundg$ in the observed range and with BBN constraints. Gaugino mass
unification relations then imply $\mgluino\lsim5.4\tev$. With
increasing $\treh$ this upper bound goes down but we still expect it
to be considerably higher than the upper limit $\mgluino\lsim1.8\tev$
claimed in ref.~\cite{fiy03} for $\treh=3\times10^9\gev$. The
difference may be due to the considerably less conservative
assumptions about the primordial abundances of light elements in
ref.~\cite{fiy03} which followed ref.~\cite{kkm04}.

\section{{\boldmath $\treh$ versus $\mgravitino$}}

We now extend our analysis to smaller $\mgravitino$ down to less than
$1\mev$. In general thermal relics with masses as low as some $10\kev$
can constitute cold DM. We note that such small values are rather
unlikely to arise within the CMSSM with the gravity--mediated SUSY
breaking scheme where $\mgravitino$ is expected to lie in the range of
several $\gev$ or a few $\tev$, as mentioned earlier. In other
SUSY breaking scenarios $\mgravitino$ can often be either much smaller
or much larger than this most natural range.  For example, in models
with gauge--mediated SUSY breaking~\cite{gudice+ratazzi99} the
gravitino can be extremely light $\mgravitino={\cal O}({\ev})$. On the
other hand, in models with anomaly--mediated SUSY
breaking~\cite{anomalysusyx} gaugino masses are typically of order
$10$ to $100\tev$.  These comments notwithstanding, in a
phenomenological analysis like this one, we therefore think it is
still instructive to display more explicitly the dependence between
$\mgravitino$ and cosmological constraints from BBN and CMB and the
ensuing implications for the maximum $\treh$.
\begin{figure}[!t]
  \begin{center}
  \begin{tabular}{c c}
   \includegraphics[width=0.5\textwidth]{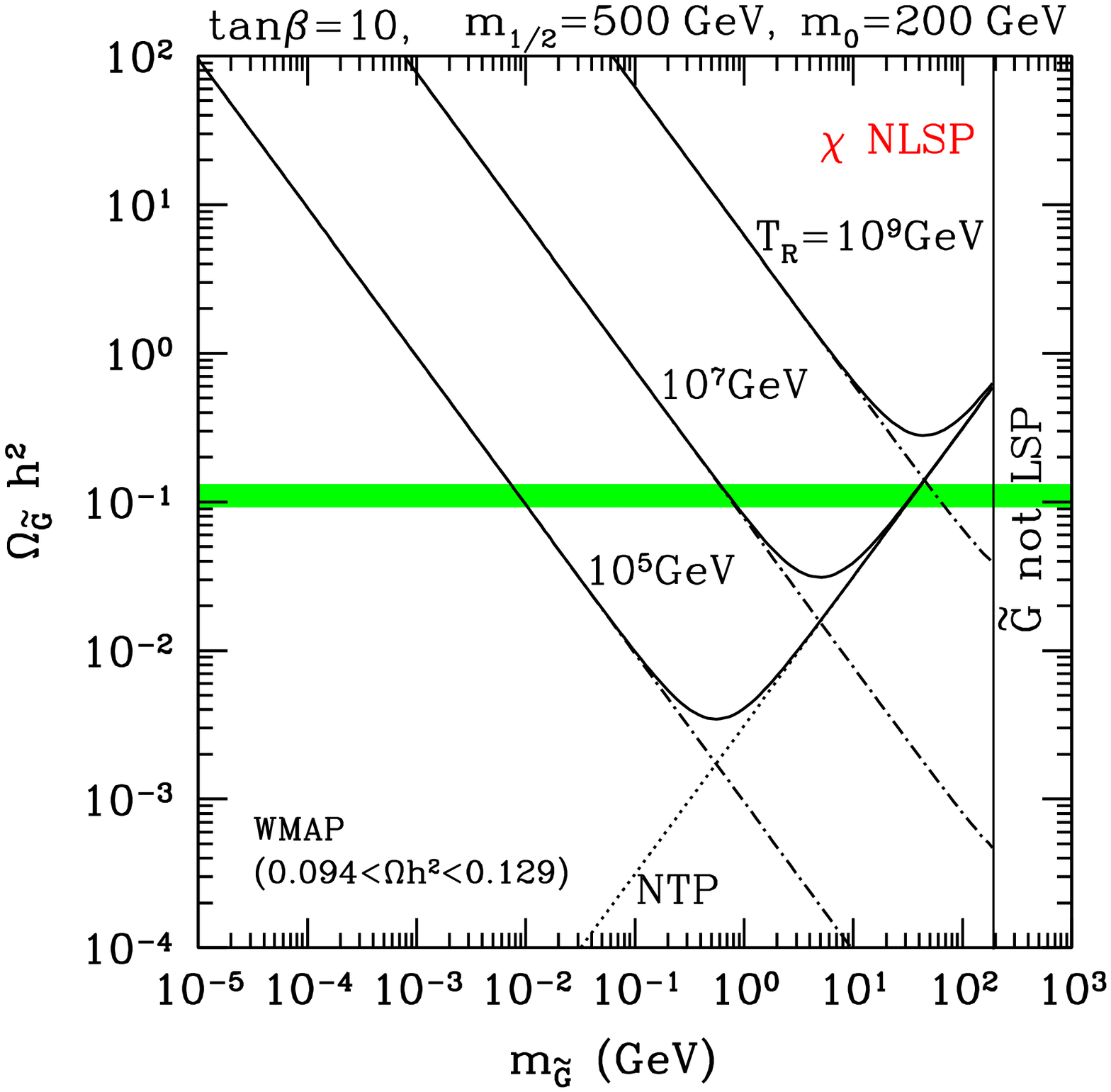}
    & 
   \includegraphics[width=0.5\textwidth]{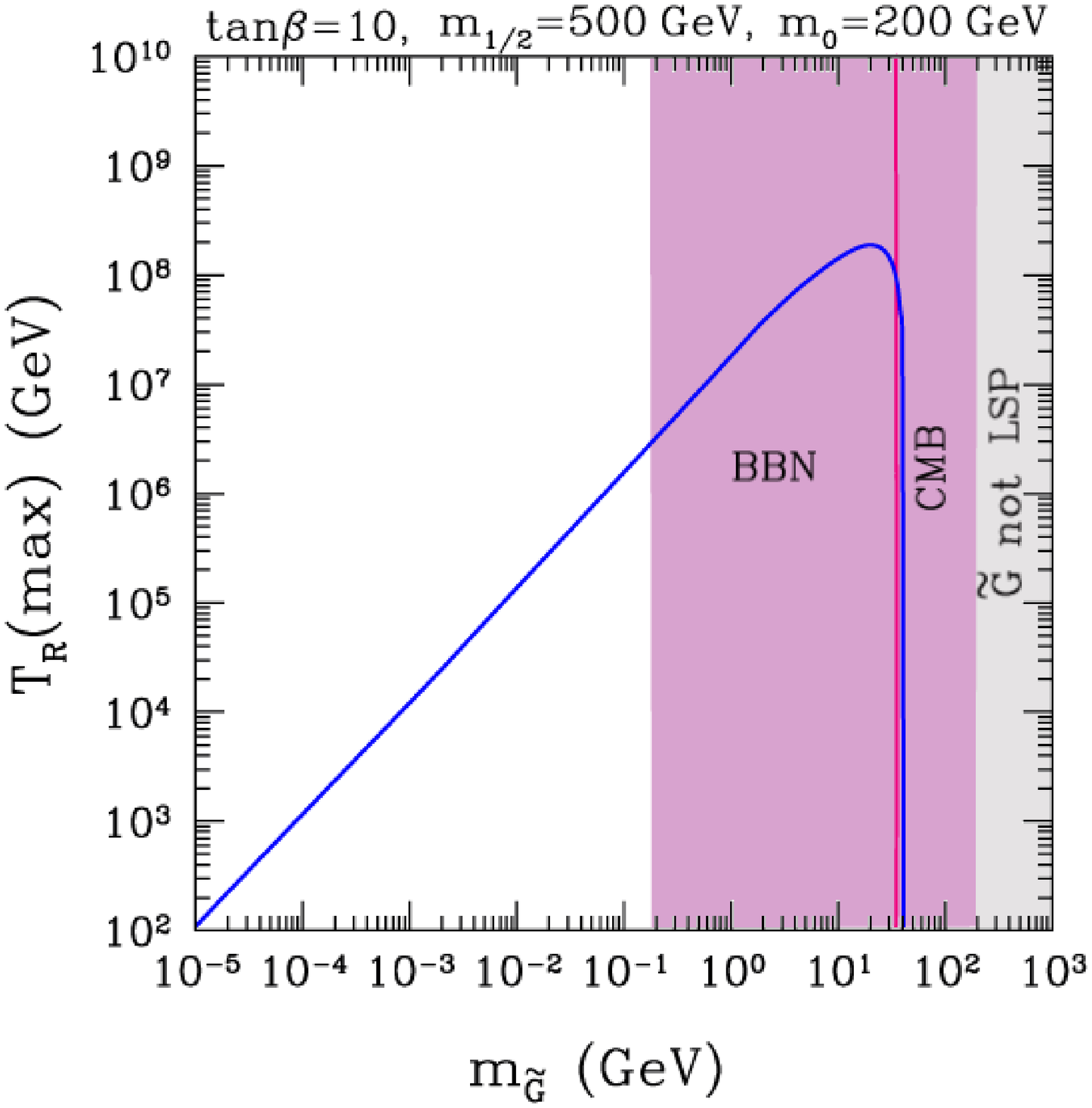}
  \end{tabular}
  \end{center}
  \caption{\small Left window: The total gravitino relic abundance
    $\abundg$ (solid lines) as a function of the gravitino mass
    $\mgravitino$ for $\tanb=10$, $\azero=0$, $\mu>0$ and for the
    point $\mhalf=500\gev$, $\mzero=200\gev$ 
($\chi$ NLSP). Thermal
    production contribution (dot--dashed lines) to $\abundg$ is shown
    for different choices of the reheating temperature ($\treh=10^9,\ 10^7,\
    10^5\gev$), while the non--thermal production one (dotted line) is
    marked by NTP. The horizontal green band shows the preferred range
    for $\abundcdm$ (marked WMAP).  Right window: The highest reheating
    temperature (blue line) versus $\mgravitino$ such that the relic
    density constraint is satisfied for the same choice of parameters
    as in the left window. The colored regions are excluded by BBN
    (violet), CMB (right side of magenta line), and the gravitino not
    being the LSP. We can see that the sub--GeV gravitino, $\treh$ as
    small as $10^5\gev$ 
are sufficient to provide the expected amount
    of DM in the Universe.}
  \label{fig:trehmaxchi}
\end{figure}
\begin{figure}[!t]
  \begin{center}
  \begin{tabular}{c c}
   \includegraphics[width=0.5\textwidth]{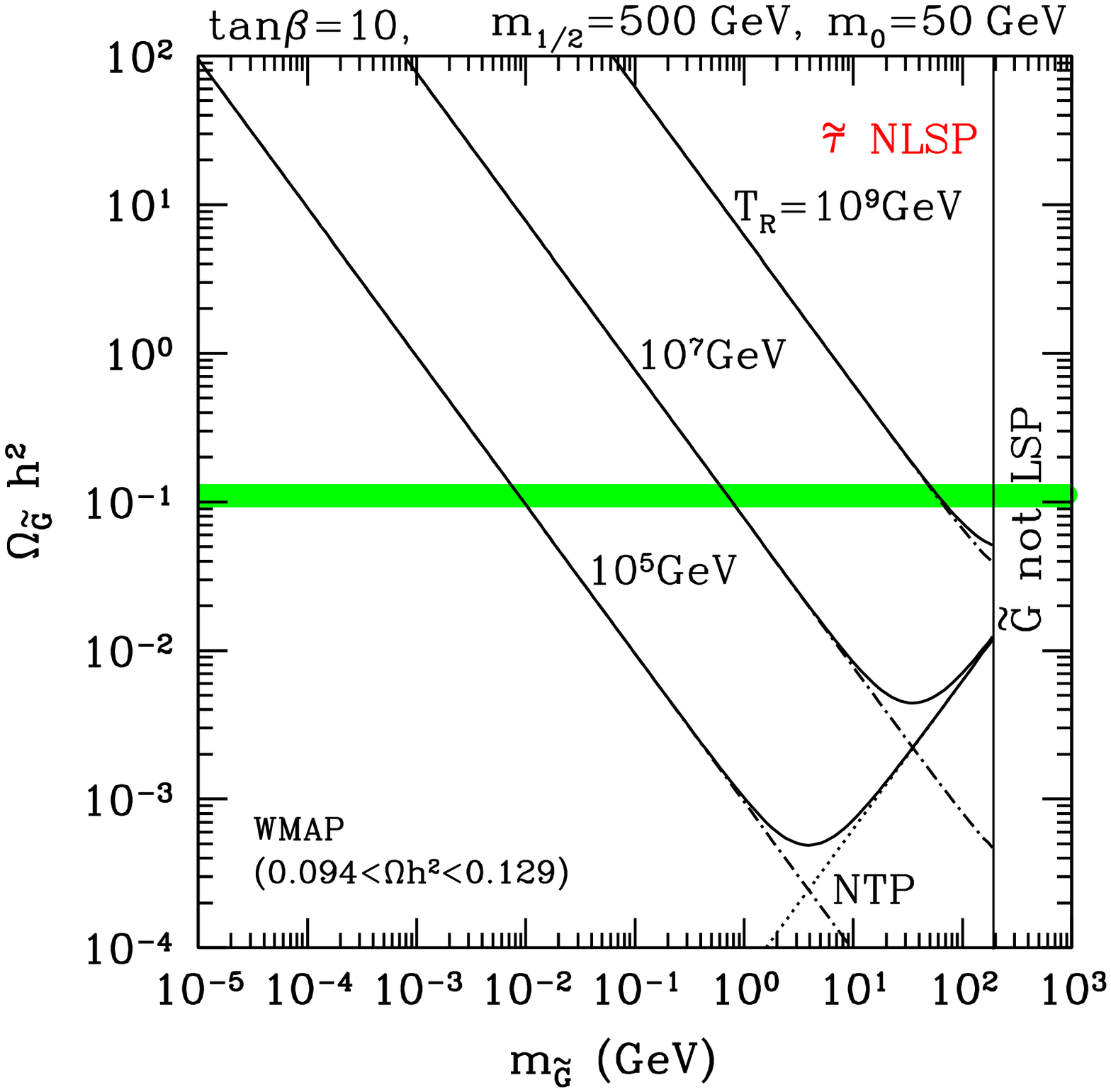}
    & 
   \includegraphics[width=0.5\textwidth]{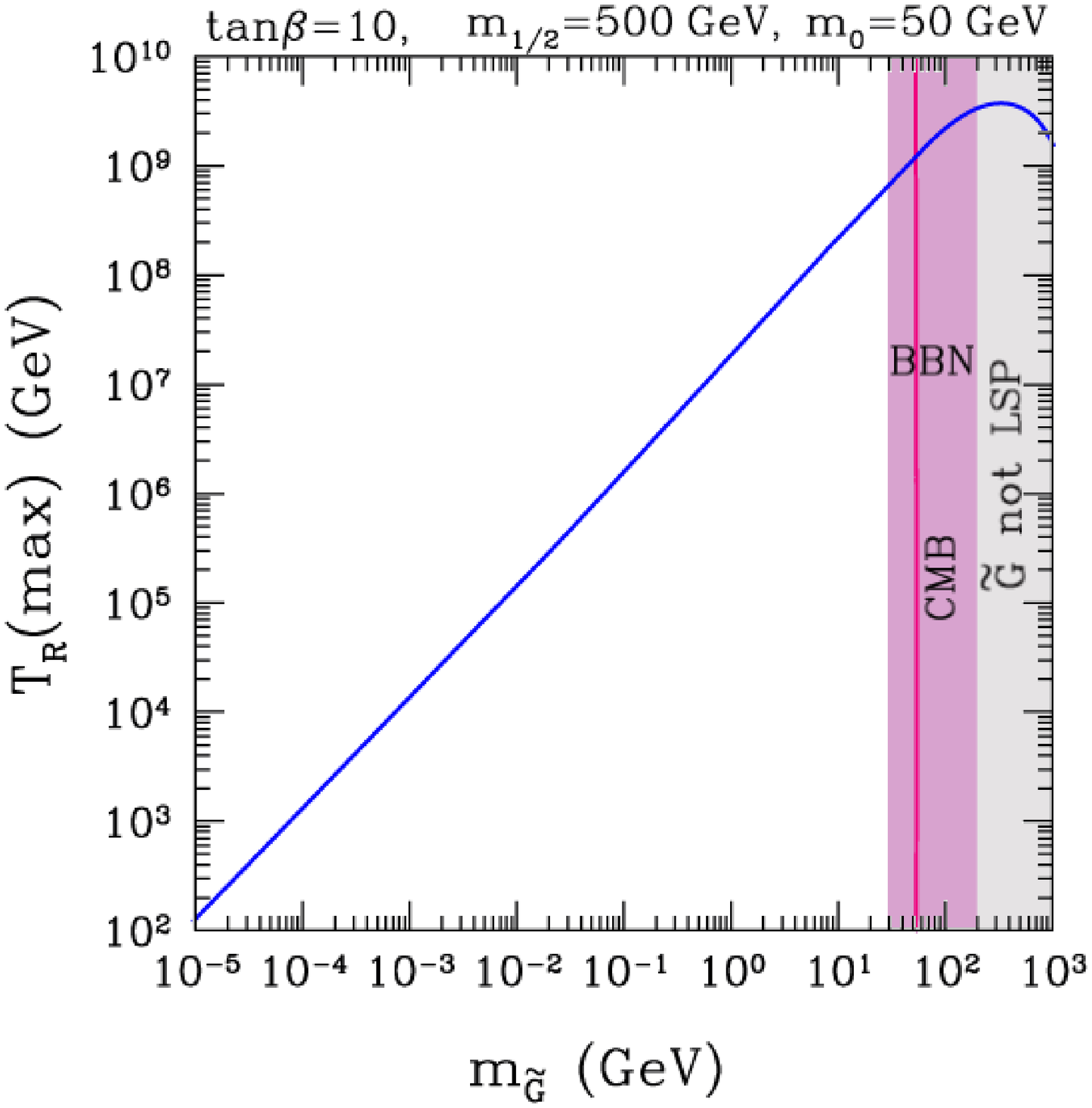}
  \end{tabular}
  \end{center}
  \caption{\small The same as fig.~\protect\ref{fig:trehmaxchi} but
  for the different point $\mhalf=500\gev$ and $\mzero=50\gev$
  ($\stauone$ NLSP).}
  \label{fig:trehmaxstau}
\end{figure}
On the other hand, in the following we will not apply the UFB
constraint. This is not only motivated by the reasons discussed above
but, more importantly, because the constraint strongly depends on the
full scalar potential which in turn depends on the field content of
the model. So far we have assumed minimal supergravity but in other
SUSY breaking scenarios several new scalars are present which are
likely to lead to very different UFB constraints.

In the left windows of figs.~\ref{fig:trehmaxchi}
and~\ref{fig:trehmaxstau} we plot the total gravitino relic abundance
$\abundg$ (solid line), its TP contribution $\abundgtp$ (dot--dashed lines) and its
NTP part $\abundgntp$ (dotted line) as a function of $\mgravitino$,
for $\tanb=10$, $\azero=0$, $\mu>0$ and $\mhalf=500\gev$ and for
several choices of $\treh$. In fig.~\ref{fig:trehmaxchi}
we take $\mzero=200\gev$ 
($\chi$ NLSP) and in fig.~\ref{fig:trehmaxstau}
$\mzero=50\gev$ ($\stauone$ NLSP). In both cases the contribution
$\abundgntp$ from NTP provides a lower limit to the total gravitino
relic abundance $\abundg$ while  $\abundgtp$ varies
with $\treh$.

In the right windows of figs.~\ref{fig:trehmaxchi}
and~\ref{fig:trehmaxstau} we plot the maximum value of $\treh$
consistent with $0.094< \abundg < 0.129$ versus $\mgravitino$ for the
same choices of other parameters as in the respective left windows. We
can see how the strong constraints from BBN and then CMB only affect
larger $\mgravitino$ in the $\gev$ range or more. Sub--$\gev$
gravitino mass leaves the CMSSM almost unconstrained by the above
constraints. In particular, the neutralino NLSP region becomes for the
most part allowed again.  Increasing $\mgravitino$ reduces the effect
of TP. This is because it becomes harder to produce them in inelastic
scatterings in the plasma. On the other hand, at some point the bounds
from BBN and CMB eventually put an upper bound on $\treh$. We 
examined a number of cases, including various
gravitino mass values but could not find consistent solutions above an
upper limit of
%
\begin{eqnarray}
\treh\lsim \textrm{a few}\times10^8\gev.
\label{trehbound}
\end{eqnarray}

No values of $\treh$ exceeding the above values were also found by
considering $\azero=\pm1\tev$, in addition to our default value of
$\azero=0$. When $\azero=1\tev$, the regions excluded by constraints
from Higgs mass bound and due to a tachyonic region become larger. In
particular, for $\tanb=50$, the Higgs mass constraint extends to
$700\gev$ and the tachyonic region increases to some $\mzero=400\gev$
and $\mhalf=1000\gev$, while the constraint due to $B\rightarrow
X_s\gamma$ becomes weaker and is burried under the Higgs mass
constraint.  When $\azero=-1\tev$, the Higgs mass constraint become
weaker but the tachyonic region become even larger and extends to
$\mhalf=1100\gev$ for $\tanb=50$.  For both values of $\azero$, the
neutralino NLSP is still diallowed, while in stau NLSP region some
allowed regions remain, similarly to the case of $\azero=0$.

By comparing the left and the right windows of
figs.~\ref{fig:trehmaxchi} and~\ref{fig:trehmaxstau}, we can again see
that the NTP contribution alone is normally not sufficient to provide
the expected range $0.094< \abundg < 0.129$. On the other hand,
assuming $\mgravitino$ as small as $100\kev$, for the gravitino to
provide most of {\em cold} DM in the Universe, implies that $\treh$ as
small as $10^3\gev$ 
is sufficient for TP to provide the
expected amount of cold DM in the Universe.

Finally, we comment on the interesting possibility that gravitino
relics may not be all cold but that a fraction of them may have been
warm at the time of decoupling.  This is because in the case of relics
like gravitinos produced in thermal processes at high $\treh$, their
momenta exhibit a thermal phase--space distribution while gravitinos
from NLSP freeze--out and decay have a non--thermal distribution. As a
result, even though both populations are initially relativistic, they
red--shift and become non--relativistic at different times and may
have a different impact on early growth of large structures, CMB and
other observable properties of the Universe.  This ``dual nature'' of
gravitinos was investigated early on in the framework of
gauge--mediated SUSY breaking scenario (in which gravitinos are light,
in the $\kev$ range)~\cite{bmy96} where the thermal population was
warm while the non--thermal one was providing a ``volatile'' component
characterized by a high {\em rms} velocity $v_{\rm rms}$. The scheme
was originally explored in~\cite{pierpaolietal95} in the case of light
axinos. Depending on the axino mass and other properties, they can provide
a dual warm--hot distribution~\cite{bgm94} or warm--cold
one~\cite{ckrplus}.  More recently, it was
found~\cite{kamionkowskietal04} that late charged (stau) NLSP decays
could suppress the DM power spectrum if NLSP decays contributed some
${\cal O}(10\%)$ of the total DM abundance and
$\taux\sim10^7\sec$. Comparing with our results we conclude that the
effect is probably marginal in the CMSSM. A similar effect was also
studied in the case of supergravity~\cite{cfrt05} and in a more
general setting in neutral heavy particle
decays~\cite{kaplinghat05}. Large $v_{\rm rms}$ may lead to the
damping of linear power spectrum and reducing the density of cuspy
substructure and concentration of halos, which have been considered to
be potential problems for the standard CDM scenario. A contribution of
such a warm (or hot) component of dark matter may, however, be
strongly limited by current and future constraints from early
reionization~\cite{jlm05}.  In the model studied here gravitinos from
TP would provide a cold component while those from NTP could be a warm
one. A more detailed investigation would be required to assess
constraints on, and implications of, this mixed warm--cold relic
gravitino population in the presented framework.

\section{Summary}

We have re--examined the gravitino as cold dark matter in the Universe
in the framework of the CMSSM. In contrast to other studies, we have
included both their thermal population from scatterings in an
expanding plasma at high temperatures and a non--thermal one from NLSP
freeze--out and decay. In addition to the usual collider constraints,
we have applied bounds from the shape of the CMB spectrum and, more
importantly, from light elements produced during BBN. The
implementation of the last constraint has in the present study been
considerably improved and also updated ranges of light element
abundances have been used but basically confirm and strengthen our
previous conclusions~\cite{rrc04}. The neutralino NLSP region is not
viable, while in large parts of the stau NLSP domain the total
gravitino relic abundance is consistent with the currently favored
range. Unless one allows for very large $\mhalf\gsim5\tev$,
for this to happen a substantial contribution from TP is
required which implies a lower limit on $\treh$.  For example,
assuming heavy enough gravitinos (as in the gravity--mediated SUSY
breaking scheme), $\mgravitino>1\gev$ leads to $\treh>10^7\gev$ 
(if allowed by BBN and other constraints). In a more generic case, if
$\mgravitino>100\kev$ then $\treh> 10^3\gev$. 
Generally, for light gravitinos ($\mgravitino\lsim1\gev$) BBN and CMB constraints
become irrelevant because of NLSP decays taking place much earlier.
On the other hand, the above constraints imply an upper bound~(\ref{trehbound}),
which appears too low for 
thermal leptogenesis, as already concluded in~\cite{rrc04}.

Finally, we have shown that in most of the stau NLSP region consistent
with BBN and CMB constraints the usual Fermi vacuum is not the global
minimum of the model. Instead, the true vacuum, while located far away
from the Fermi vacuum is color and charge breaking. 

Implications for SUSY searches at the LHC are striking. The standard
missing energy and missing momentum signature of a stable neutralino
LSP is not allowed in this model, unless $\mgravitino\lsim
1\gev$. Instead, the characteristic signature would be a detection of
a massive, (meta--)stable and electrically charged particle (the
stau). It is worth remembering that such a measurement would not be a
smoking gun for the gravitino dark matter since in the case of the
axino as CDM the stau NLSP (as well as neutralino NLSP) is typically
allowed as well~\cite{ckrplus}. Finally in some cases it may be possible
to accumulate enough staus to be able to observe their decays into
gravitinos~\cite{bhrey04} and/or to distinguish them from decays into
axinos~\cite{bchrs05}. SUSY searches at the LHC open up a realistic
possibility of pointing at non--standard CDM candidates and
additionally revealing the vacuum structure of the Universe.

\acknowledgments D.G.C. and K.--Y.C. are funded by
PPARC. L.R. acknowledges partial support from the EU Network
MRTN-CT-2004-503369.  We would further like to thank the European
Network of Theoretical Astroparticle Physics (ENTApP, part of ILIAS,
contract number RII3-CT-2004-506222) for financial support.  \\

\noindent{\bf \Large Erratum}

Following a recent paper of Pradler and Steffen~\cite{ps06} a formula
for the thermal production of gravitino in ref.~\cite{bbb00} (Bolz, et
al.) has been corrected. In addition, we have corrected a numerical
error in our routine computing $\alpha_s$ at high temperatures. As a
consequence, the regions of $\abundg$ (green bands) in all the figures
have shifted to the left, towards smaller $\mhalf$ relative to the
previous version, which in turn has led to improving the upper bound
on the reheating temperature by about an order of magnitude. The new
bound is $T_R \lesssim \textrm{a few} \times 10^8 \gev$.

We thank J.~Pradler and F.~Steffen for checking our results and
informing us about a discrepancy with theirs.

\end{document}